\documentclass[traditabstract]{aa}  

\usepackage{graphicx}
\usepackage{txfonts}
\usepackage{lscape}

\graphicspath{{figures/}{/Users/rgb/Downloads/astro/ml/pdf/}}

\usepackage{hyperref}

\hypersetup{breaklinks=true,colorlinks=true,linkcolor=blue,citecolor=blue,urlcolor=blue}

\usepackage{natbib,twoopt}
\bibpunct{(}{)}{;}{a}{}{,}             

\defcitealias{dr3}{DR3}
\defcitealias{Bell:2003}{B03}
\defcitealias{Bruzual:2003}{BC03}

\newcommand\starlight{{\sc starlight}}                  

\newcommand{\hlr}{HLR} 
\newcommand{\hmr}{HMR} 

\newcommand{\mustar}{$\Sigma_{\star}$} 

\newcommand{\ngal}{452}
\newcommand{\oimf}{0.29}

\newcommand{\gi}{$\bigtriangledown_{in}$}
\newcommand{\go}{$\bigtriangledown_{out}$}

\newcommand{\ml}{M/L}
\newcommand{\cml}{color-M/L}
\newcommand{\lmlc}{log(M/L)-color}

\newcommand{\mlb}[1]{$\rm M/L_{#1}$}
\newcommand{\mll}{M/L$_{\lambda}$}
\newcommand{\mlld}{M/L$^{Dered}_{\lambda}$}
\newcommand{\mlbc}[3]{${\rm M/L_{#1}}-(#2-#3)$}
\newcommand{\lmlbc}[3]{${\rm log(M/L_{#1}})-(#2-#3)$}

\newcommand{\band}[1]{$#1$ band}

\newcommand{\wpycasso}{\url{http://pycasso.iaa.es/ML}}

\newcommand{\wcb}{\url{http://www.bruzual.org/~gbruzual/cb07}}

\newcommand{\pycasso}{{\sc p}y{\sc casso}}
\newcommand{\stl}{{\scshape starlight}}

\newcommand{\suf}{_px1.q057.d22a512.ps03.k1.mE.CCM.Bzca6e}

\begin{document} 

\titlerunning{ML}

\title{Spatially resolved mass-to-light from the CALIFA survey} 
\subtitle{Mass-to-light ratio vs. color relations}

\author{
        R.~Garc\'ia-Benito\inst{\ref{inst1}}
        \and R.~M.~Gonz\'alez Delgado\inst{\ref{inst1}}
        \and E.~P\'erez\inst{\ref{inst1}}
        \and R.~Cid Fernandes\inst{\ref{inst2}}
        \and S.~F.~S\'anchez\inst{\ref{inst3}}
        \and A.~L.~de~Amorim\inst{\ref{inst2}}
}

\institute{Instituto de Astrof\'isica de Andaluc\'ia (CSIC), P.O. Box 3004, 18080 Granada, Spain\label{inst1} 
(\email{rgb@iaa.es})
   \and Departamento de F\'isica, Universidade Federal de Santa Catarina, P.O. Box 476, 88040-900, Florian\'opolis, SC, Brazil\label{inst2}
   \and Instituto de Astronom\'ia, Universidad Nacional Aut\'onoma de M\'exico, Circuito Exterior, Ciudad Universitaria, Ciudad de M\'exico 04510, M\'exico\label{inst3}
}

\date{}

\abstract{
We investigated the mass-to-light versus color relations (MLCRs) derived from the spatially resolved star 
formation history (SFH) of a sample of \ngal\ galaxies observed with integral field spectroscopy in the 
CALIFA survey. We derived the stellar mass ($M_\star$) and the stellar mass surface density (\mustar) 
from the combination of full spectral fitting (using different sets of stellar population models) with 
observed and synthetic colors in optical broad bands. This method allows obtaining the radial structure of 
the mass-to-light ratio ($M/L$) at several wavelengths and studying the spatially resolved MLCRs. Our 
sample covers a wide range of Hubble types from Sc to E, with stellar masses ranging from 
$M_\star \sim 10^{8.4}$ to $10^{12}$ M$_\odot$. The scatter in the MLCRs was studied as a function of 
morphology, stellar extinction, and emission line contribution to the colors. The effects of the initial mass function
(IMF) and stellar population models in the MLCRs were also explored. Our main results are that
{\em (a)} the $M/L$ ratio has a negative radial gradient that is steeper within the central 1 half-light-radius 
(HLR). It is steeper in Sb-Sbc than in early-type galaxies.
{\em (b)} The MLCRs between $M/L$ and optical colors were derived with a scatter of $\sim$ 0.1 dex. 
The smallest dispersion was found for the combinations ($i$, $g-r$) and ($R$, $B-R$).
Extinction and emission line contributions do not affect the scatter of these relations. Morphology 
does not produce a significant effect, except if the general relation is used for galaxies redder than 
$(u-i) > 4$ or bluer than $(u-i) < 0$.
{\em (c)} The IMF has a large effect on MLCRs, as expected. The change from a Chabrier to a Salpeter IMF produces
a median shift of $\sim$ \oimf\ dex when mass loss from stellar evolution is also taken into account. 
{\em (d)} These MLCRs are in agreement with previous results, in particular for relations with $g$ and $r$ 
bands and the $B$ and $V$ Johnson systems. 
}

 


\maketitle
%

\section{Introduction}

During the past decade, large-scale  surveys of galaxies such as the Sloan Digital Sky 
Survey (SDSS, \citet{SDSS02}) or the Galaxy And Mass Assembly (GAMA, \citet{Driver:2011}) 
survey, have shown that sorting galaxies by their stellar mass ($M_\star$) is a useful way 
to classify them. This is so because the stellar mass is strongly correlated with other 
global galaxy properties such as the stellar mass surface density (\mustar) 
\citep{Kauffmann:2003, Kauffmann:2003a}, age, and metallicity of the stellar populations 
\citep{Gallazzi:2005,  CidFernandes:2005, Gallazzi:2006, Mateus:2006, Asari:2007}, and with the
star formation rate (SFR) \citep{Brinchmann:2004}. 
The correlations and the scatter of these global relations can be understood as a sequence on mass 
where the position of each galaxy is a consequence of their mass growth assembly history and 
its evolutionary state (e.g., \citealt{Behroozi:2013}).

Other works based on spatially resolved data have shown that \mustar\ is a fundamental 
parameter that drives\footnote{Both $M_\star$ and \mustar\ are the primary product of the SFH 
in galaxies and should be considered as proxies of a more fundamental parameter, the gravitational 
potential (local or global).} the star formation history (SFH) of galaxies \citep{Bell:2000}. 
More recently, integral field spectroscopic surveys\footnote{CALIFA \citep{Sanchez:2012}, 
SAMI \citep{Croom:2012}, or MaNGA \citep{Bundy:2015}.} have found local relations between 
\mustar\ and local stellar ($Z_\star$, \citealt{GonzalezDelgado:2014b}) and gas metallicity 
\citep{Sanchez:2013}, the age of the stellar populations \citep{GonzalezDelgado:2014a, Goddard:2016, Scott:2017}, 
and the star formation rate \citep{GonzalezDelgado:2016, CanoDiaz:2016}. These local relations are 
similar to the global ones, implying that \mustar\ regulates the star formation in disks, 
while $M_\star$ drives the star formation in the spheroidal components.

$M_\star$ and \mustar\ are thus important properties of galaxies that cannot be measured directly. Deriving 
these quantities from observed data involves stellar populations synthesis (SPS) models. The relation 
between light and mass can be obtained by modeling the following three properties: 

\begin{itemize}
\item The galaxy spectrum by fitting its stellar continuum with a combination of single stellar population models 
(e.g., \citealt{Panter:2003, CidFernandes:2005, Tojeiro:2011, Perez:2013, Sanchez:2016, Garcia-Benito:2017}), 
or selected spectral indices with a library of models of parametric SFHs 
(e.g., \citealt{Kauffmann:2003, Gallazzi:2005, LopezFernandez:2016, LopezFernandez:2018}).

\item The galaxy spectral energy distribution from optical-to-near-IR (NIR) broadband photometry, usually with a 
library of parametric SFHs (e.g.,\ \citealt{Taylor:2011}; see \citealt{Walcher:2011} and 
\citet{Conroy:2013} for reviews).

\item The relation between colors and the mass-to-light ratio at some wavelength:

\begin{equation}
\label{Eq:MLCR}
 \log {M/L_{\lambda_i}} = a_{\lambda_i} + b_{\lambda_i} \times (m_{\lambda_j} - m_{\lambda_k}), 
 \end{equation}
 
\noindent where the bands $\lambda_i$, $\lambda_j$, and $\lambda_k$ may be independent, or 
$\lambda_i$ = $\lambda_j$ or $\lambda_k$ 
\citep{Bell:2001, Bell:2003, Zibetti:2009, Gallazzi:2009, Taylor:2011, Into:2013, McGaugh:2014, Roediger:2015}.

\end{itemize}

Certainly, the $\rm{M/L_\lambda}$-color relation (MLCR) is the simplest method to derive $M_\star$ 
(and \mustar), as it relies on photometry in only two bands. However, the uncertainty in mass is usually 
larger with MLCR than with the other two methods, and it depends on the SFH, the initial mass 
function (IMF), and the extinction assumed to model the MLCR \citep{Conroy:2013}. 

In their pioneering work, \citet{Bell:2001} found that along the SFH (parametrized by a $\tau$ model), 
galaxies move across a well-defined locus in the space of $M/L_B$ versus $B-R$, suggesting that the $B-R$ 
color is a good proxy for $M/L_B$. In a more recent work, using an extensive library of SFHs and the 
stellar masses derived by SED fitting of the SDSS bands for the GAMA survey, \citet{Taylor:2011} found 
that their calibration $M/L_i$ versus $(g-i)$ color can be used to estimate $M_\star$ with an accuracy 
of $\leq$ 0.1 dex. Another important conclusion from their work is that the age-metallicity-dust 
degeneracy helps in the estimation of $M/L_i$ because it moves the galaxies along the 
$M/L_i$ versus $(g-i)$ relation.

From stellar population synthesis models, the amplitude of the evolution of $M/L$  is much less 
pronounced in the NIR than at optical wavelengths \citep{Starburst99, Bruzual:2003}. Recent works 
based on Spitzer data use a constant value of $\sim0.5$ M$_\odot$/L$_\odot$ to convert the 3.6$\mu$m 
emission into $M_\star$ \citep{McGaugh:2014}. Other studies found that $M/L_{3.6 \mu m}$ 
also depends on the $[3.6]-[4.5]$ color \citep{Meidt:2012}, and this color may be contaminated 
by non-stellar sources \citep{Querejeta:2015}. Furthermore, modeling the asymptotic giant branch 
phase of stellar evolution is crucial in population synthesis models \citep{Into:2013}, and it may 
have a major impact on $M/L$ at NIR wavelengths.

Most previous works to date have obtained MLCR based on the integrated $M/L_\lambda$ 
\citep{Bell:2003, Gallazzi:2009, Taylor:2011, Into:2013, Roediger:2015}. However, galaxies have 
$M/L_\lambda$ and color gradients, and the effects of spatial variations of the SFH, extinction, 
metallicity, and age on the MLCR have not been explored before. This is the goal of this work. 
We use the full spectral synthesis technique by fitting the spatially resolved optical 
spectroscopy provided by the CALIFA survey to obtain the spatially resolved $M/L$, retrieving 
for each spaxel in the galaxy the SFH (as in \citealt{GonzalezDelgado:2017}), the stellar extinction, 
metallicity, and ages (as in \citealt{GonzalezDelgado:2015}), the recent SFR 
(as in \citealt{GonzalezDelgado:2016}), and \mustar\ (as in \citealt{GonzalezDelgado:2014a}). 
Optical colors are measured on the observed spectra and on the synthetic ones to explore 
the effect from the emission lines on the colors in the MLCR. Because the CALIFA sample covers 
all Hubble types, we are able to explore the radial profiles of $M/L_\lambda$ and their gradient 
with galaxy morphology, and their effect on the MLCR.

This paper is organized as follows. Section \ref{Sec:sampledata} describes the observations and 
the properties of the galaxies analyzed here. Section \ref{Sec:Method} explains the analysis method
for retrieving the spatially resolved SFH and \mustar, and how colors are measured. Section \ref{Sec:ML} 
presents the radial profiles of $M/L$. In Section \ref{Sec:MLCRs} we derive the MLCRs and compare 
these results with those from the literature. Section \ref{Sec:Summary} reviews our main findings. 

Throughout this work we assume a flat $\Lambda$CDM cosmology with $\Omega_{M}$ = 0.3, 
$\Omega_{\Lambda}$ = 0.7, and H$_{0}$ = 70 km s$^{-1}$ Mpc$^{-1}$. 


\section{Sample and data \label{Sec:sampledata}}

\subsection{Sample \label{subsec:sample}}

The sample was selected from the final CALIFA data release \citep[][hereafter DR3]{dr3}, with a 
total 646 galaxies observed with the V500 grating, 484 with the V1200 grating, and 446 galaxies with the 
COMB setup (the combination of the cubes from both setups). The DR3 release is the combination of two 
samples: a) the CALIFA main sample (MS) consists of galaxies belonging to the CALIFA mother sample, with 
a total of 529 observed with V500 and 396 with the COMB, all included in the DR3; b) the rest of the 
galaxies belong to the CALIFA extension sample, a set of galaxies observed within the CALIFA 
collaboration as part of different ancillary science projects (see \citetalias{dr3}).

The CALIFA mother sample is fully described in \citet{Walcher:2014}. The main properties of this 
sample are 
{\em (a)} an angular isophotal diameter between $45{\tt''}$  and $79{\tt''}$;  
{\em (b)} a redshift range $0.005 \leq z \leq 0.03 $; and  
{\em (c)}  color ($u-r < 5$) and magnitude ($-24 < M_r < -17$), thus covering the whole color-magnitude 
diagram. The sample is not limited in volume, but it can be "volume-corrected", allowing us to provide 
estimates of the stellar mass function \citep{Walcher:2014} and other cosmological observables such 
as $\rho_{\rm SFR}$ \citep{GonzalezDelgado:2016, LopezFernandez:2016, LopezFernandez:2018} and $\rho_\star$.

The mother and extended CALIFA samples were morphological classified through visual inspection of the 
SDSS $r$-band images. As in our previous works (e.g., \citealt{GonzalezDelgado:2016, Garcia-Benito:2017}), 
we grouped the galaxies into seven morphological bins: 
E (65 galaxies), S0 (54, including S0 and S0a), Sa (70, including Sa and Sab), Sb (75), 
Sbc (76), Sc (77, including Sc and Scd), and Sd (35, including Sd, Sm, and Irr).

\subsection{Data}
\label{subsec:data}

Observations were carried out at the 3.5m telescope at Calar Alto observatory with the Postdam 
Multi Aperture Spectrograph \citep[PMAS,][]{Roth:2005} in the PPaK mode \citep{Verheijen:2004}, 
created for the Disk Mass Survey project \citep{Bershady:2010}. PPak contains a bundle of 331 science 
fibers of 2.7\arcsec\ diameter each and a 71\arcsec $\times$ 64\arcsec\ field of view 
\citep[FoV;][]{Kelz:2006}. 
The observations were planed to observe each galaxy with two different overlapping setups. 
Here, we analyze only the low-resolution setup (COMB; R$\sim$850) 
that covers from 3745 \AA\ to 7500 \AA\ with a spectral resolution of 
$\sim$ 6 \AA\ full width at half-maximum (FWHM). The spatial sampling of the data cubes is 
1\arcsec/spaxel with a point spread function (PSF) of FWHM $\sim$ 2.6\arcsec, which at the 
mean redshift of the sample corresponds to a physical resolution of $\sim$ 0.7 kpc. The data 
were calibrated with version V2.2 of the reduction pipeline \citep{dr3}. Details about the 
observational strategy and data processing are given in 
\citet{Sanchez:2012}, \citet{dr1}, \citet{dr2}, and \citet{dr3}.


\section{Method: 2D maps of stellar mass, colors, and \mll \label{Sec:Method}}

\begin{figure*}
\centering
\resizebox{0.49\textwidth}{!}{\includegraphics{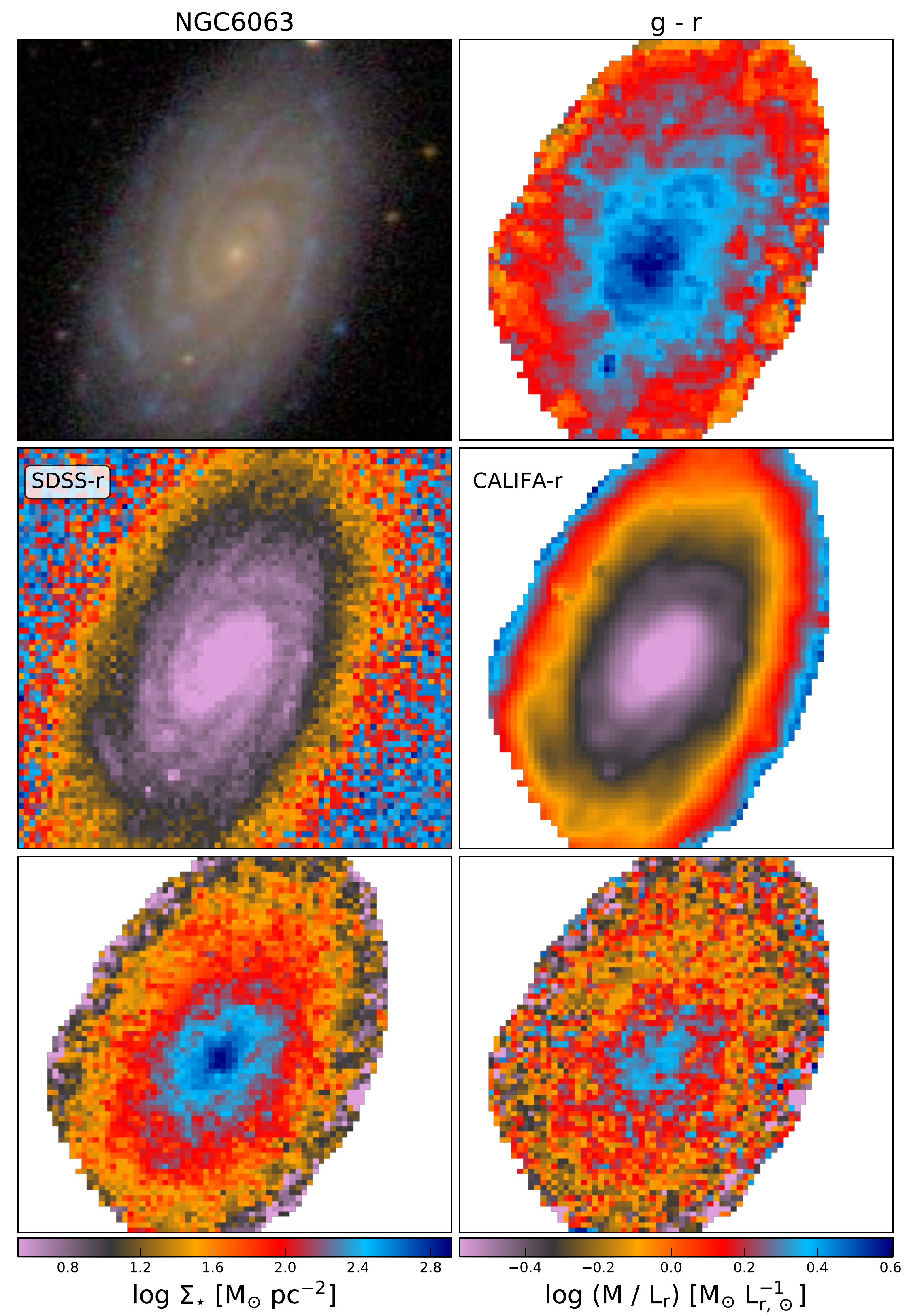}} 
\resizebox{0.49\textwidth}{!}{\includegraphics{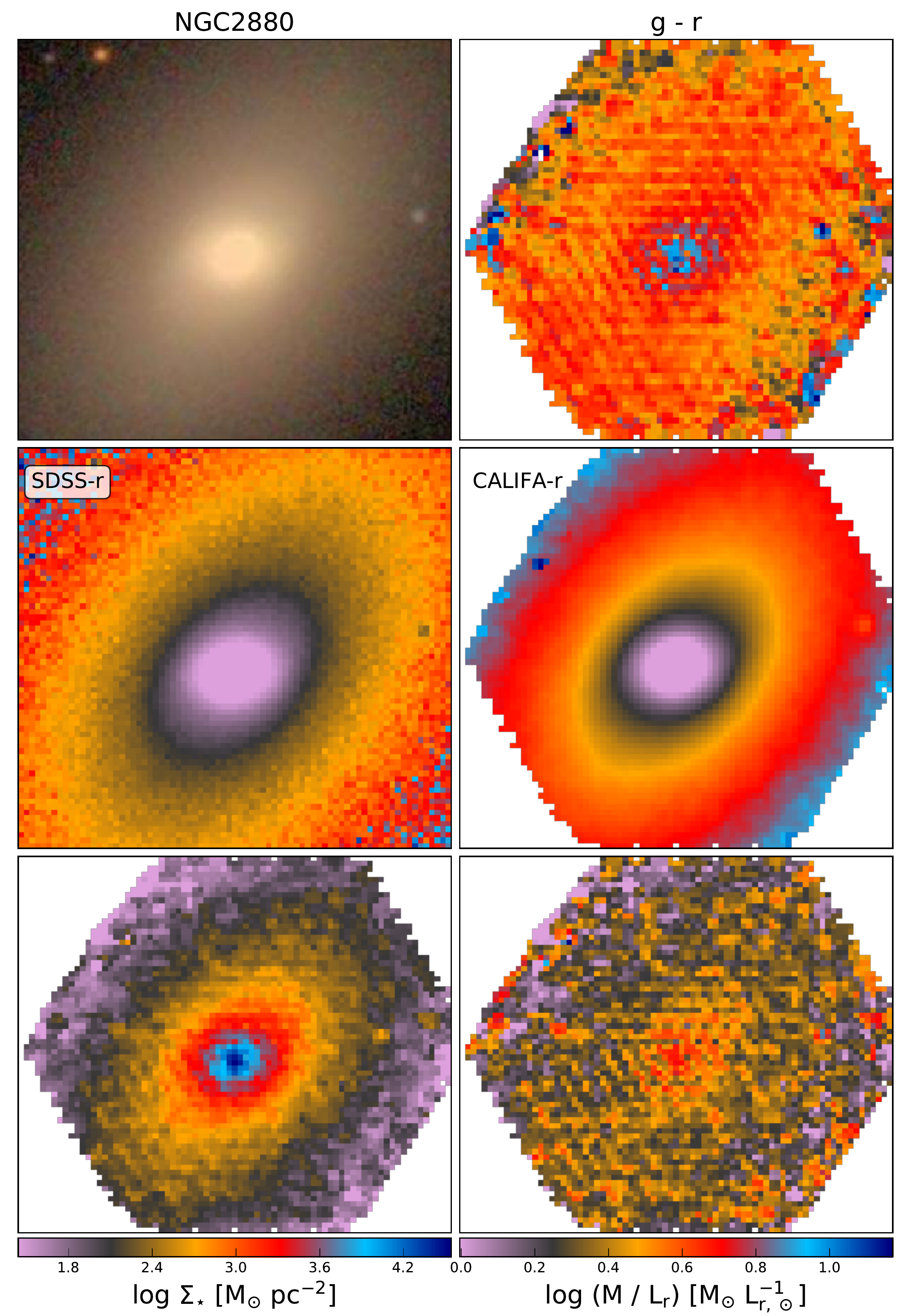}} 
\caption{Examples of some of the data products used in the analysis of this work.
2D maps obtained from the CALIFA data cubes for NGC 6063 (unbarred Sbc, left) 
and NGC 2880 (E7, right). For each galaxy, the figure shows maps of $g-r$ (upper right), 
$r$ band (middle right), \mustar\ (lower left), and \mlb{r} (lower right). For the sake 
of comparison, a three-color postage-stamp SDSS image (upper left) and an SDSS $r$ band 
(middle left) reprojected to the same pixel scale as the CALIFA data cubes are also 
shown.
}
\label{fig:2Dmaps}
\end{figure*}

\subsection{Spatially resolved SFH}

For this work, we obtained the spatially resolved SFH of each galaxy to derive the stellar mass 
surface density (\mustar) and $M_\star$. We followed the same method as in previous works 
(e.g., \citealt{Perez:2013}, \citealt{GonzalezDelgado:2014a}, \citealt{GonzalezDelgado:2015}). 
In short, we fit with \starlight\ \citep{CidFernandes:2005} the spectrum of each individual 
spaxel (pixelwise) within the isophote level where the average signal-to-noise ratio (S/N) 
$\geq$ 3, decomposing the spectra in terms of stellar populations with different ages and 
metallicities. 

We used base {\rm CBe}, a set of 246 SSPs from an updated version of 
\citet[][hereafter BC03]{Bruzual:2003} models (Charlot \& Bruzual 2007, hereafter CB07; 
private communication\footnote{\wcb}). 
In CB07, the spectral library STELIB \citep{Leborgne:2003} is replaced by a combination of 
the MILES \citep{Sanchez-Blazquez:2006, Falcon-Barroso:2011} and {\sc granada} \citep{Martins:2005} 
libraries. The evolutionary tracks are those collectively known as Padova 1994 
\citep{Alongi:1993, Bressan:1993, Fagotto:1994a, Fagotto:1994b, Girardi:1996}. 
The metallicity covers $\log Z_\star/Z_\odot = -2.3$, $-1.7$, $-0.7$, $-0.4$, 0, and $+0.4$, 
while ages run from 1 Myr to 14 Gyr. The IMF is that of \citet{Chabrier:2003}. Dust effects were 
modeled as a foreground screen with a \citet{Cardelli:1989} reddening law with R$_V$ = 3.1.

MILES and STELIB differ in a number of ways. MILES has a larger number of stars and 
a wider range in stellar parameters 
\citep[effective temperature, gravity, and metallicity; see][]{Sanchez-Blazquez:2006}. 
While both MILES and STELIB have only few massive stars, in our analysis MILES is complemented 
with the synthetic stellar library GRANADA \citep{Martins:2005} and the SSP templates by 
\citet{GonzalezDelgado:2005}.
Although the stellar metallicity in galaxies is a parameter that always has a large uncertainty, 
several works indicate that results are more consistent for the same data fitted with MILES + GRANADA 
than with STELIB; for example, for CALIFA data sets \citep[e.g. Fig.9 in][]{CidFernandes:2014}
or for LMC and SMC stellar clusters \citep{GonzalezDelgado:2010}.

The results were then processed through \pycasso\ (the Python CALIFA \starlight\ Synthesis Organizer; 
\citealt{CidFernandes:2013}; \citealt{Amorim:2017}) to produce a suite of spatially resolved stellar 
population properties. \pycasso\ organizes the information into a multi-dimensional data 
structure with spatial as well as age and metallicity dimensions. From these, 2D maps of stellar mass 
surface density, \mustar, stellar extinction (A$_V$), and luminosity surface density were 
obtained to derive 2D maps and radial profiles of {\rm M/L}.

\subsection{2D maps of stellar mass}
\label{Sec:mstar}

For each spectrum, the stellar mass was derived from \starlight\ as explained in \citet{CidFernandes:2013}. 
To reflect the mass currently in stars, the initial mass formed in stars is corrected for the mass 
returned to the interstellar medium during stellar evolution. These results were then processed through 
\pycasso\ to produce 2D maps of the stellar mass distribution. The galaxy's total stellar mass was obtained 
by adding the masses of individual spaxels. This method takes into account the spatial variations of 
the SFH and extinction across the face of the galaxy. We also take into account areas that were masked 
in the data cube, replacing the missing spaxels by the average values at the same radial distance. 

Masses obtained with base {\rm CBe} are on average \oimf\ dex lower than those used in 
\citet{Garcia-Benito:2017}. This is the factor expected due to change of the IMF from Chabrier 
to Salpeter.
Throughout this paper all the stellar masses used in the \ml\ and MLCR (either in 2D maps 
or profiles) come from these \starlight\ results, unless explicitly stated otherwise.

\subsection{Optical images and colors\label{Sec:colors}}

We computed colors in two alternative ways: 1) the CALIFA data cubes were convolved with the SDSS 
$g$ and $r$ and the Johnson $B$ and $V$ filter responses. The label \textit{Obs} was attached to 
results that were computed using these data; 2) synthetic data cubes were computed by assigning to 
each spaxel the corresponding synthetic spectrum derived from the full spectral fit. These synthetic 
data cubes were convolved with all the filters available within the wavelength range, generating 
images in the $u, g, r, i, z$ SDSS bands and the $B$, $V$, $R$ Johnson bands. The label \textit{Syn} 
was attached to results that were computed using these data.

The two methods allow us to estimate the variations of the MLCRs that are due to the effect of stellar 
extinction and the contribution from emission lines. Furthermore, when the synthetic spectra are used, the 
second method provides images in optical bands at wavelengths that are not covered by our observations. 
We derived both restframe and observed colors for each galaxy. However, because our galaxies are 
in a very small range of redshift, the difference is very small, $\sim$ 0.01 magnitudes. In what 
follows, all results are calculated in restframe unless noted otherwise.

Since \starlight\ provides the stellar extinction, the luminosity (and colors) in the \ml\ 
(and therefore in the MLCR) can be given as not dereddened (\mll) or as dereddened quantities (\mlld). 
In this work, all luminosity and color values are given uncorrected for reddening, unless otherwise 
specified. Obviously, the mass is the same in both cases.

\subsection{2D maps of \mll \label{Sec:mlmaps}}

For each galaxy, the image of \mll\ was obtained by dividing the 2D map of stellar mass by the image 
in $L_\lambda$, where $\lambda$ can be any of the optical SDSS bands or the $BVR$ Johnson bands. 
Figure \ref{fig:2Dmaps} shows examples of the images obtained for two different galaxies. To the left,  
NGC 6063: $g-r$ color obtained with method (2) and ${\rm M/L_r}$. The SDSS three-color postage-stamp 
image is also shown. NGC 6063 (CALIFA 823) is an isolated unbarred Sbc galaxy \citep{Walcher:2014}. 
The stellar mass map shows a clear radial variation from the inner to the outer parts of the galaxy, 
with a more pronounced gradient in the inner \hlr\ and smoother profile in the outer one. This is 
also reflected in the \mlb{r} map. This behavior is typical of Sbc galaxies 
(see Fig. \ref{fig:mlradial}).

Figure \ref{fig:2Dmaps} (right) shows the example for NGC 2880 (CALIFA 272), an isolated E7 galaxy. The 
color distribution is more homogeneous in this case, as expected for this type of galaxy. The mass map 
gradient is clearly more pronounced. The \mlb{r} map is smoother than for NGC 6063, but a 
gradient is still present, as seen in Fig. \ref{fig:mlradial}.

2D maps of $M/L_\lambda$ for each galaxy, together with the average radial profiles and MLCRs (see 
section \ref{Sec:MLCRs}), are available in \wpycasso. Our magnitudes in the SDSS bands 
are given in the $AB$ system \citep{Oke:1983}, while the Johnson $BVR$ are  Vega magnitudes. 
The $M/L_\lambda$ is given in solar units, and it can be transformed from $M_\odot$/$L_\odot$ to 
$M_\odot$/ erg s$^{-1}$ Hz$^{-1}$ by adding 2.05, 1.90, and 1.81 to $\log {\rm M/L_\lambda}$ 
for $\lambda$ = $g$, $r$, and $i$, respectively.


\section{Radial structure of the mass-to-light ratio}
\label{Sec:ML}

This section presents a series of results derived from the spatially resolved $M/L_\lambda$, 
including the radial structure of $M/L_\lambda$ and the radial gradients as a function of the 
galaxy morphology and galaxy mass.
For each galaxy, the radial variation of $M/L_\lambda$ was obtained by compressing each individual 
2D map in azimuthally averaged radial profiles. As in our previous studies 
(e.g., \citealt{GonzalezDelgado:2014a, GonzalezDelgado:2015, Garcia-Benito:2017}), the radial 
distance is expressed in units of the galaxy's half-light-radius (\hlr), a convenient metric 
when averaging radial information for different galaxies. The \hlr\ is defined as the semi-major 
length of the elliptical aperture in the data cube that contains 50$\%$ of the luminosity at 
5635 \AA. To obtain the radial profile, we used elliptical apertures of thickness  0.1 \hlr. 
The ellipticity and position angle were derived from the moments of the 5635 \AA\ flux image 
of the data cube. Similarly using the mass, we define the half-mass-radius (\hmr) as the 
semi-major length of the elliptical aperture that contains 50$\%$ of the mass.

\subsection{Radial profiles\label{sec:mlradial}}

\begin{figure*}
\centering
\resizebox{\hsize}{!}{\includegraphics{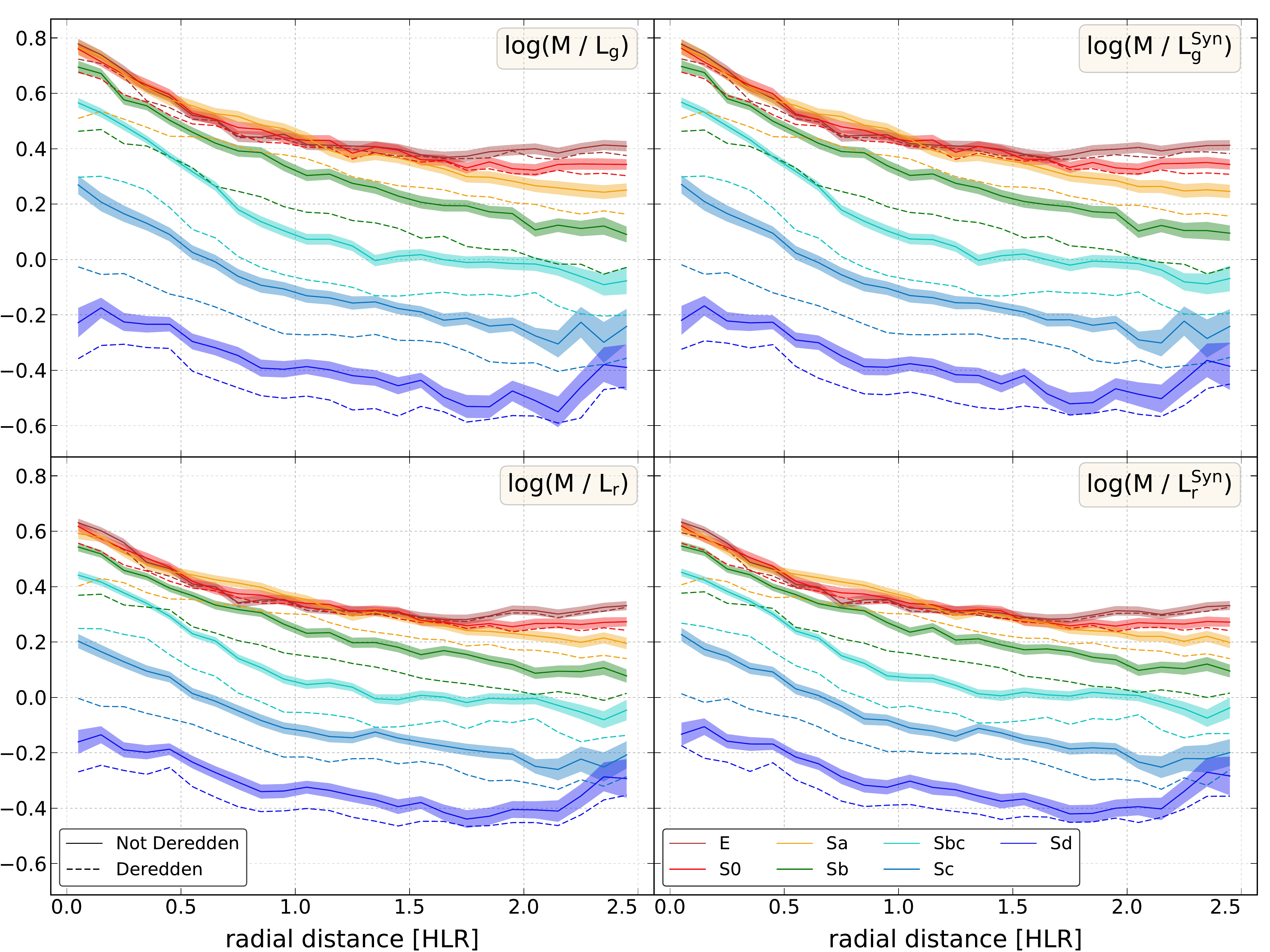}}
\caption{Radial profile of \ml\ for \band{g} (upper panels) and \band{r} (lower panels) stacked by Hubble 
type for the observed (left panels) and synthetic restframe spectra (right panels). The profiles with 
intrinsic extinction are shown in continuous lines and the deredden profiles (using the extinction values 
given by the stellar population analysis) in dashed lines. For the sake of clarity, the uncertainty band 
is plotted only in the not dereddened profiles.}
\label{fig:mlradial}
\end{figure*}

Figure \ref{fig:mlradial} shows azimuthally averaged radial profiles of \mlb{g} and \mlb{r}. Results 
are stacked by Hubble type in seven morphological classes. In the left panels, the profiles are obtained 
from the observed spectrum, both not\textup{} dereddened (continuous lines) and dereddened (dashed lines), 
using the results from the stellar population fitting. In the right panels we use the \mll\ images obtained 
from the synthetic spectra, also corrected and not corrected for stellar extinction.

All profiles decrease outward, with inner regions having higher \ml. The profiles scale with the Hubble 
type. At any given distance, \ml\ is higher for early-type galaxies than for late-type spirals. Taking as 
reference the value at 1 \hlr, the logarithmic value of \mlb{g} (\mlb{r}) ranges from $-0.39$ ($-0.32$) 
to 0.42 (0.32) for Sd to E galaxies. 
The effect of the extinction is clearly seen in Fig. \ref{fig:mlradial}. This effect is more 
significant in the central regions of intermediate-type spirals, in particular in Sb/Sbc, where 
the value of the extinction and its central gradient is higher \citep{GonzalezDelgado:2015}.

The effect of emission lines  in \ml\ is very small, as can be seen from the comparison between 
observed and synthetic profiles in Fig. \ref{fig:mlradial}. For \mlb{g} (\mlb{r}), the maximum difference 
is around 0.01 dex (0.02 dex) for late-type spirals.

\begin{figure}
\centering
\resizebox{\hsize}{!}{\includegraphics{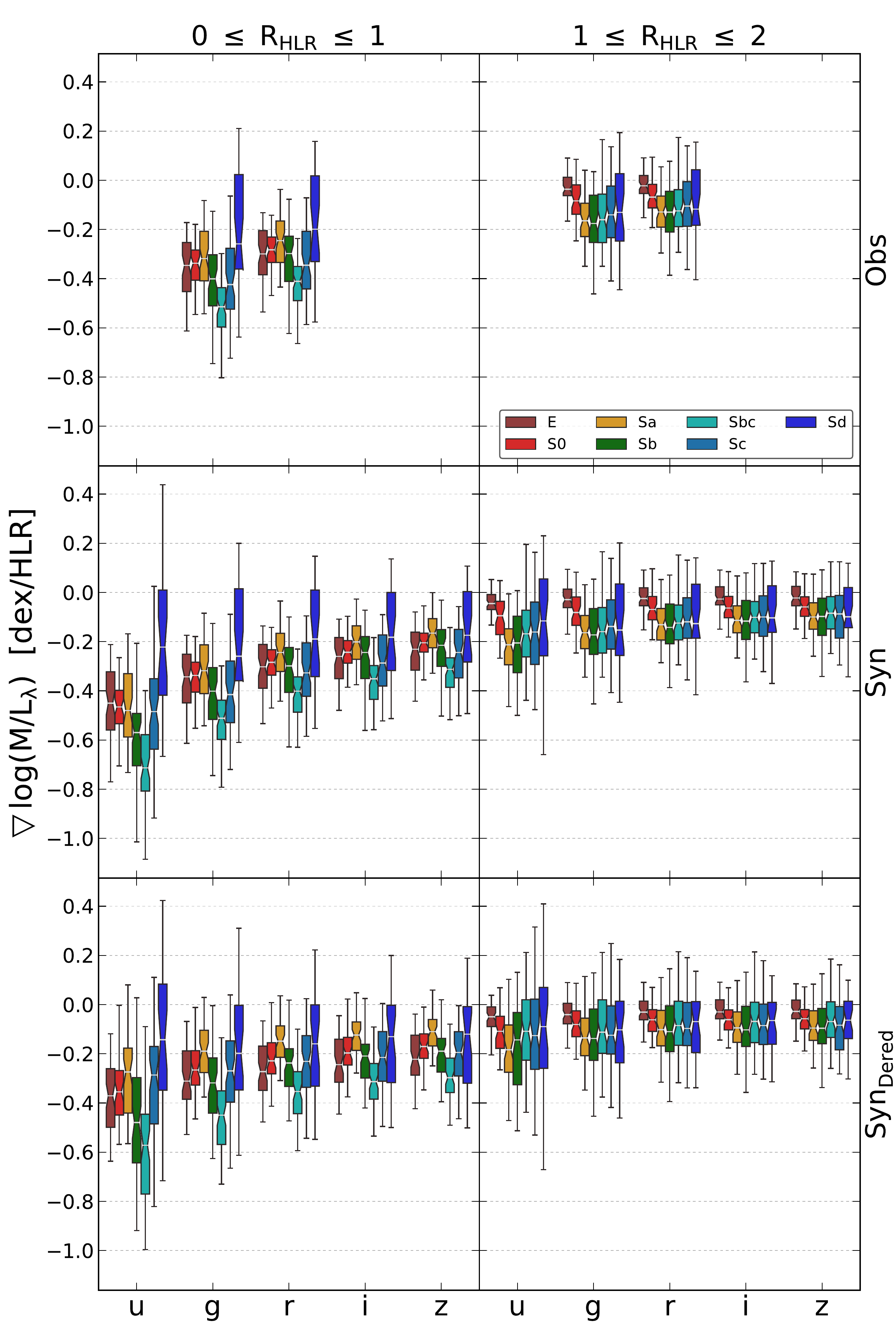}}
\caption{Box plot diagrams of the radial gradients of \mll\ measured in the inner \gi\ 
(0 $\leq$ \hlr\ $\leq$ 1; left panels) and outer \go\ (1 $\leq$ \hlr\ $\leq$ 2; right panels) 
spatial regions in the observed spectra (upper panels; \textit{Obs}), synthetic spectra 
(middle panels; \textit{Syn}), and dereddened synthetic spectra 
(lower panels; \textit{Syn$_{Dered}$}).}
\label{fig:gML}
\end{figure}

\subsection{Radial gradients}

We can analyze Fig. \ref{fig:mlradial} in another way. As in \citet{Garcia-Benito:2017}, we performed 
a robust linear fit of the log M/L$_\lambda$ profile over a radial range. We defined an inner radial 
gradient as the slope of the robust linear fit over the entire inner 1 \hlr\ (\gi) and in a similar 
way to the outer (1 $\leq$ \hlr $\leq$ 2) radial gradient (\go). Expressed in this way, the gradients 
have units of dex/\hlr. 

Figure \ref{fig:gML} presents box plot diagrams of the radial gradients of \mll\ in these two spatial 
regions, measured in the observed (upper panels; $\rm M/L^{Obs}$), in the synthetic spectra 
(middle panels; $\rm M/L^{Syn}$), and in the dereddened syntetic spectra (lower panels; 
$\rm M/L^{Syn}_{Dered}$) for the SDSS bands. All of them show negative gradients, although 
their values are higher in the inner than in the outer regions. As mentioned in Sect. \ref{sec:mlradial}, 
the difference between synthetic and observed profiles is very small, which is also reflected in their 
gradients (for the filters in common). 

There are two clear trends. The first is seen in the values of the gradient for a given filter as 
function of Hubble type. For the inner regions, the median value of the gradients of early-type galaxies 
(E, S0, Sa) displays a similar value. The gradient for spiral galaxies (Sb, Sba, Sc) decreases 
(more negative) to the minimum, shown by Sbc galaxies. Finally, the gradient rises again up to Sd 
galaxies, which have less steep slopes and the highest dispersion in their values, mimicking the 
HMR/HLR relation found in \cite{GonzalezDelgado:2015} and \cite{Garcia-Benito:2017}. The scatter is 
similar for early and intermediate types, but it is larger for late types. The behavior is 
slightly different for the outer regions. The slope is almost flat for E galaxies, with more negative 
values up to Sa galaxies, whereas the median value of the slope is almost constant for late-type 
galaxies. 

The second trend can be appreciated by comparing the gradients obtained with different filters. In the 
inner regions, blue filters have more negative slopes than red filters, while in the outer regions 
the values are almost the same, with nearly no dependence on the filter. The scatter is in general 
lower for the outer regions, and blue filters (particularly $u$ and $g$) present a larger scatter 
than the remaining filters. In general, the outer regions show flatter (gradients) profiles than the 
inner regions.

It is worth noting that the gradients estimated using dereddened luminosities ($Syn_{Dered}$) 
present the same relative trends as the not dereddened values ($Syn$). The only difference lies 
in the absolute values, which show slightly flattened gradients for the dereddened luminosities.



\section{Spatially resolved MLCRs \label{Sec:MLCRs}}

Several factors determine the position of a galaxy on the \cml\ diagram: metallicity, reddening, 
and SFH. In addition, these properties are expected to vary from place to place within a 
particular object and in different degrees depending on the type of galaxy. If the MLCRs are 
calibrated based on models, these have to take into account all the possible variations of the 
variables involved. The same argument can be applied to observational data: the underlying sample 
has to cover the full parameter space. We remark that the CALIFA sample selection allows the 
MLCRs to be calibrated with an homogeneous and complete sample of galaxies \textit{\textup{with}} spatially 
resolved data.

Since our methodology provides the radial profile of any magnitude (mass, 
luminosity, magnitudes, colors, etc.) as described in Sect. \ref{Sec:ML}, 
it is fairly straightforward to compute the linear relations of \lmlc\ quantities. We limit the 
spatial extension of our fits to 0 $\leq$ R$\rm{_{HLR}}$ $\leq$ 2. Unlike many integrated 
relations found in the literature, we are computing spatially resolved MLCRs.
The CALIFA dataset allows us to compute MLCRs for the whole sample and 
for different subsets, grouped by Hubble types.

Tables \ref{tb:AVTABLE_SynR_AB_px1}, \ref{tb:AVTABLE_SynR_Mag_px1}, 
\ref{tb:hub3_SynR_AB_px1}, \ref{tb:hub3_SynR_Mag_px1}, 
\ref{tb:hubtyp_SynR_AB_px1}, and \ref{tb:hubtyp_SynR_Mag_px1} 
present the slope, origin (following equation \ref{Eq:MLCR}), and scatter of 
the linear \lmlc\ relations for each filter combination, both 
for SDSS ($ugriz$; Tables \ref{tb:AVTABLE_SynR_AB_px1}, \ref{tb:hub3_SynR_AB_px1}, 
\ref{tb:hubtyp_SynR_AB_px1}) and Johnson-Cousin bands ($BVR$; 
Tables \ref{tb:AVTABLE_SynR_Mag_px1}, \ref{tb:hub3_SynR_Mag_px1}, \ref{tb:hubtyp_SynR_Mag_px1}). 
All are computed in restframe in the not dereddened synthetic\textup{} spectra  
with base {\rm CBe} and Chabrier IMF. The monochromatic mass-to-light ratios $(M/L_{\lambda})$ 
are in solar units. The SDSS $ugriz$ filters are in the AB magnitude system and the Johnson-Cousins 
$BVR$ filters are in the Vega magnitude system. $\sigma_{\lambda}$ is the scatter of the residuals 
of the relation \lmlc.

We have computed the MLCRs for different types of 
samples and subsamples: 1) the whole sample comprising all \ngal\ galaxies 
(Tables \ref{tb:AVTABLE_SynR_AB_px1}, \ref{tb:AVTABLE_SynR_Mag_px1}); 2) subsamples
of early- (E, S0, Sa), intermediate- (Sb, Sbc), and late-type (Sc, Sd) galaxies 
(Tables \ref{tb:hub3_SynR_AB_px1}, and \ref{tb:hub3_SynR_Mag_px1}); and 3) grouped by 
Hubble type (Tables \ref{tb:hubtyp_SynR_AB_px1} and \ref{tb:hubtyp_SynR_Mag_px1}).

\begin{figure}
\centering
\resizebox{0.49\textwidth}{!}{\includegraphics{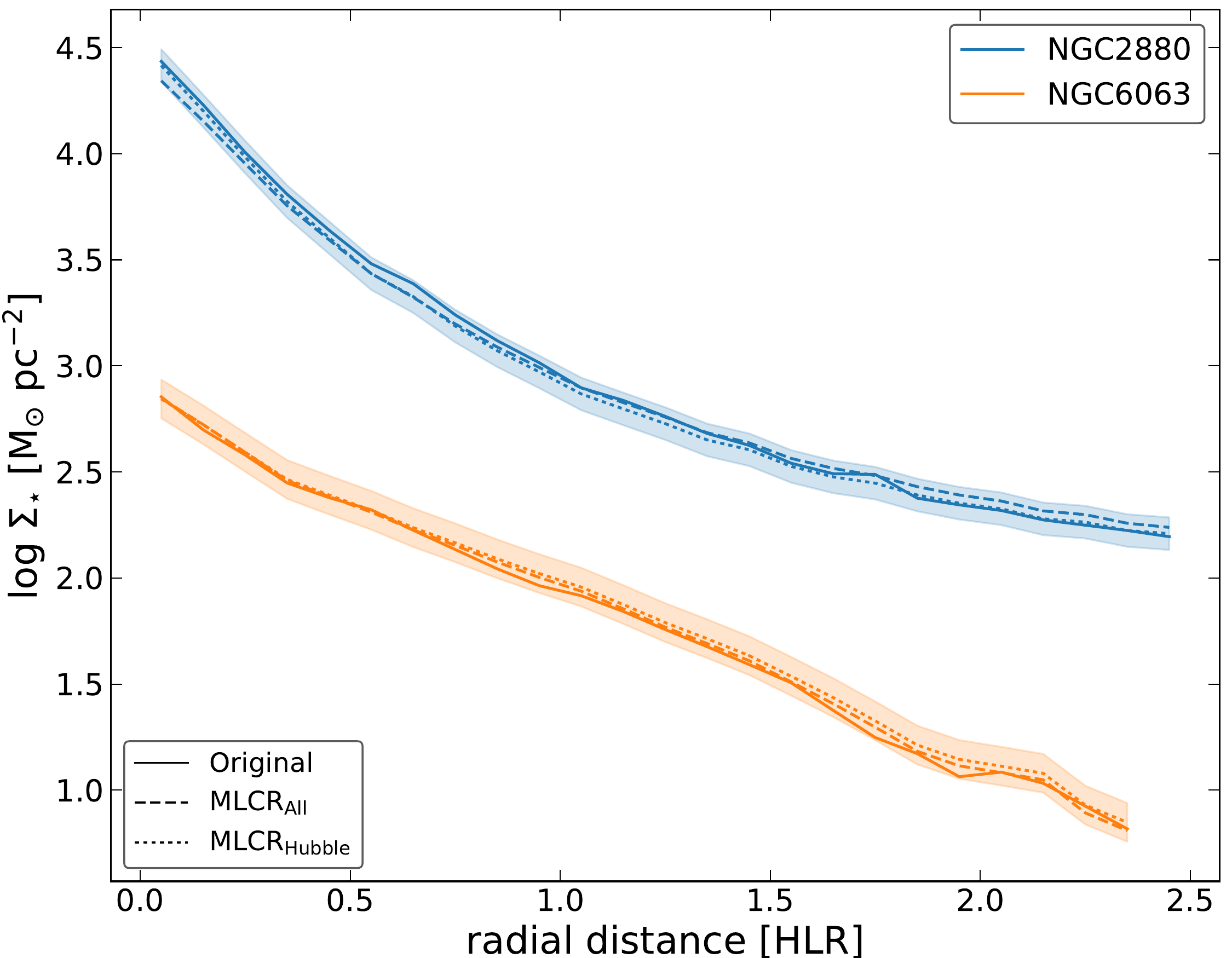}}
\caption{Comparison of radial profiles of \mustar: the original profile estimated 
from the stellar population results (continuous line), the profile derived from our MLCRs 
calibrated using all galaxies (dashed line), and the MLCRs by Hubble type (dotted line), 
for NGC 6063 and NGC 2880 (same galaxies as in Fig. \ref{fig:2Dmaps}).
The \mustar\ profiles derived from the MLCRs have been obtained from the $g-r$ 2D color 
map and the L$_g$ 2D map. For the sake of clarity, only the uncertainty band from the 
Hubble type MLCR has been plotted (the others are on the same order).}
\label{fig:mcorsd}
\end{figure}

Since these relations have been computed using spatially resolved data, it is possible 
to reproduce profiles just by using colors maps. Figure \ref{fig:mcorsd} compares the 
\mustar\ radial profile obtained from the stellar population results (continuous line) 
and those derived using our MLCRs for galaxies NGC 6063 and NGC 2880 (same objects as 
in Fig. \ref{fig:2Dmaps}). The latter profiles have been computed from the $g-r$ 2D color 
and L$_g$ maps using the MLCR calibrated using all galaxies (dashed line) and the Hubble-type relations (dotted line). It is clear from the plot that the profiles are very close 
to the originals within the errors. It is worth saying that these relations should be 
used only for evolved galaxies (galaxies at low redshift). At high redshift the trends
might depart from the correlations since the age-metallicity relation could be 
different because the stellar populations would be younger. \citet{Garcia-Benito:2017}
showed that 80\% of the mass of the galaxies was already formed  more than 10 Gyr ago 
(redshift $\sim$ 1.6), so that beyond this redshift limit the correlations might not 
be valid.

Appendix \ref{Apx} includes the MLRCs using intrinsic luminosites, that is, using the dereddened \textup{}synthetic spectra. These tables may be useful for comparison with 
models or simulations\footnote{Our webpage \wpycasso\ also hosts the tables for all 
the MLCRs.}.

\subsection{Uncertainties on MLCRs}

\begin{figure}
\centering
\resizebox{\hsize}{!}{\includegraphics{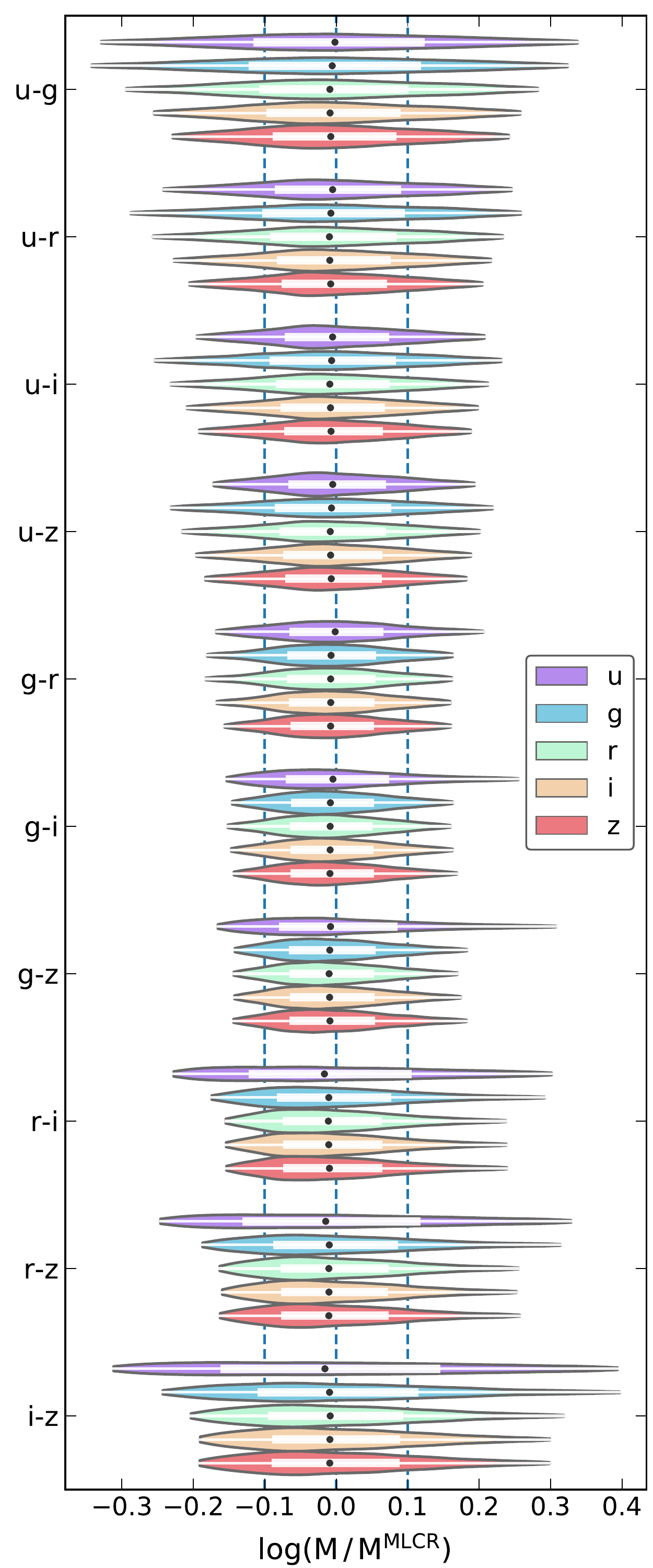}}
\caption{Violin plots of the difference between the spatially resolved mass obtained by applying our MLCRs 
and the mass derived from the stellar population fitting (Sect. \ref{Sec:mstar}) for the whole sample and 
all Sloan filter combinations. The corresponding box plot showing the interquartile range is plotted inside 
each violin plot. The inner dot in the box plot represents the median of the distribution.}
\label{fig:mass_residuals}
\end{figure}

We can assess the uncertainty on the derivation of \ml\ by estimating the scatter of 
the residuals of our $\rm{M/L_\lambda}$-color relations to the linear fit. In Tables 
\ref{tb:AVTABLE_SynR_AB_px1}, \ref{tb:AVTABLE_SynR_Mag_px1}, 
\ref{tb:hub3_SynR_AB_px1}, \ref{tb:hub3_SynR_Mag_px1}, 
\ref{tb:hubtyp_SynR_AB_px1}, and \ref{tb:hubtyp_SynR_Mag_px1} 
we present the slope, origin, and scatter of the fits for each filter combination. The 
scatter estimates the error in the final calculation of the mass. The mean scatter 
using all filter combinations is $\sim$ 0.13 dex. The minimum scatter corresponds to the 
pairs ($i$, $g-i$), ($i$, $g-r$), ($r$, $g-i$), and ($R$, $B-R$), all with a value lower than 
0.1 dex. The user can explore the tables to find the most suitable filter combination according 
to the type of galaxy. 

To check the validity of our results and the estimated values of their uncertainties, we used these 
MLCRs to compute the mass for all our spatially resolved data (radial values) for the whole sample. 
From the measured colors (\ref{Sec:colors}), we obtained the M/L ratio and then we multiplied this 
value by the corresponding luminosity value (Sect. \ref{Sec:mlmaps}). Then, we computed the logarithmic 
difference between the masses from our stellar population fittings (M$_{\star}$, Sect. \ref{Sec:mstar}) 
and the masses derived from the MLCRs (M$^{MLCR}_{\star}$). The violin plots of the distribution of 
these residuals are shown in Fig. \ref{fig:mass_residuals} for all Sloan filter combinations. 
The standard deviation of these distributions is an estimate of the uncertainties of the mass 
from the MLCRs. The values are almost equal to the scatter given in our tables. As expected, the 
distribution of the differences are centered around zero and it is easy to spot the combinations with 
less scatter.

\subsection{Effect of extinction}

\begin{figure}
\centering
\resizebox{\hsize}{!}{\includegraphics{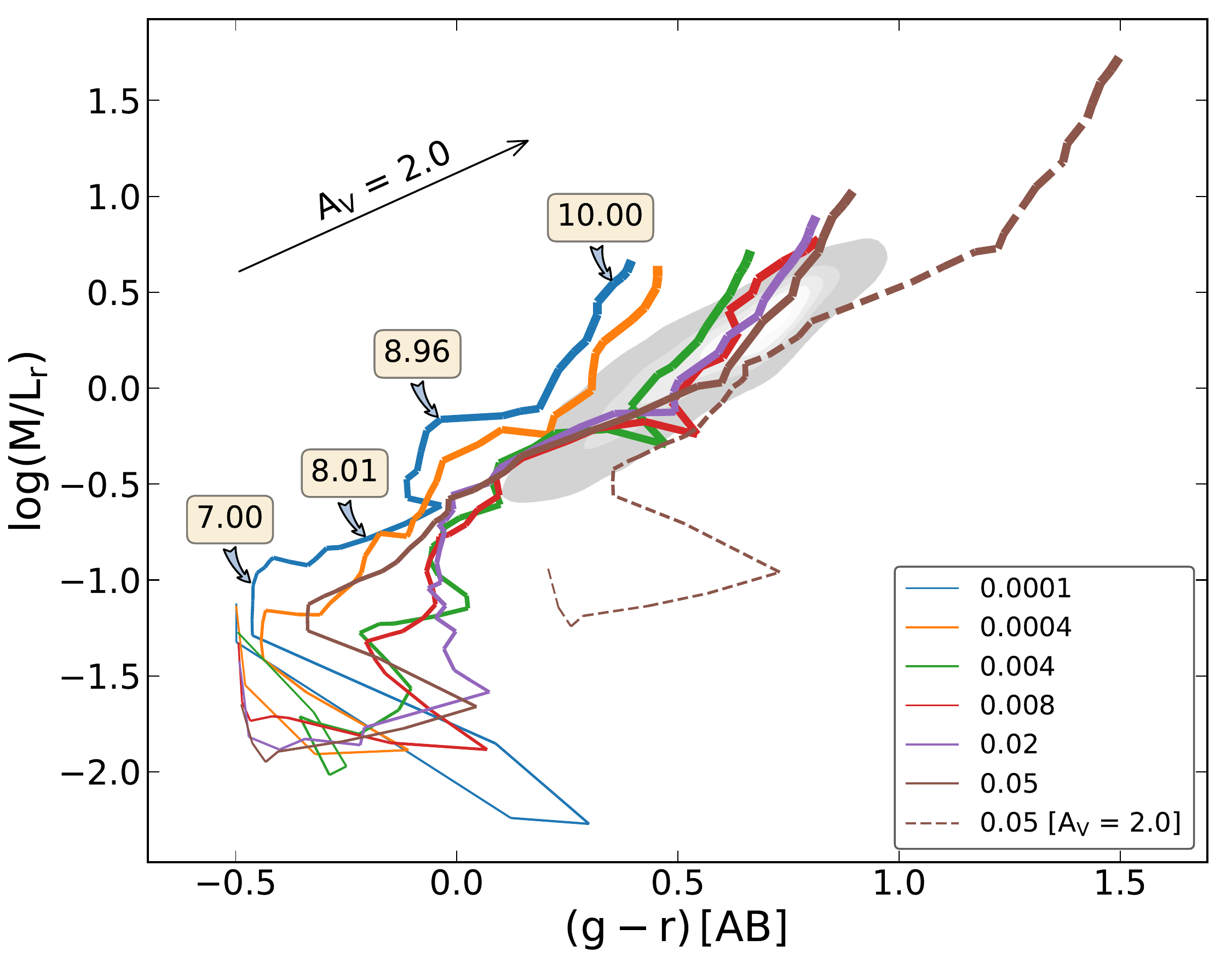}}
\caption{Tracks of the SSPs of the {\rm CBe} base in the \lmlbc{r}{g}{r} plane. The library contains 6 
metallicities and 41 ages. The tracks increase their line width with age (the older the age, 
the thicker the line). Four signposts mark the position and value of the (logarithmic) age in the 
lowest metallicity track. The highest metallicity track reddened by A$_V$ = 2 mag together with a
reddening vector of this amount are also shown. The density distribution of our sample is underlaid 
in gray contours encompassing 95\% of the points.}
\label{fig:MLCR_SSPs}
\end{figure}

In Fig. \ref{fig:mlradial} the effect of the extinction in the radial profiles is clearly appreciated.
We examine now the behavior of extinction in the \lmlc\ plane. To better understand the position of 
galaxies in this  space, we have computed the color ($g-r$) and \mlb{r} of all SSPs of our 
{\rm CBe} base. Figure \ref{fig:MLCR_SSPs} shows the path of equal metallicity SSPs with age (the 
thicker the line, the older the age). We also show our sample distribution in gray contours 
encompassing 95\% of the points. The reddening vector is also shown.
All tracks move very quickly and appear hectic during the first few million 
years. After 10 Myr the tracks stabilize and follow a smoother path with some small changes at several points. Clearly, our distribution lies in the loci of intermediate to old age and 
medium to high metallicity unextincted tracks. Of course, younger ages of the same metallicities or 
lower metallicities can fall in the distribution area with different amounts of extinction, 
following the reddening vector.

\begin{figure*}
\centering
\resizebox{\hsize}{!}{\includegraphics{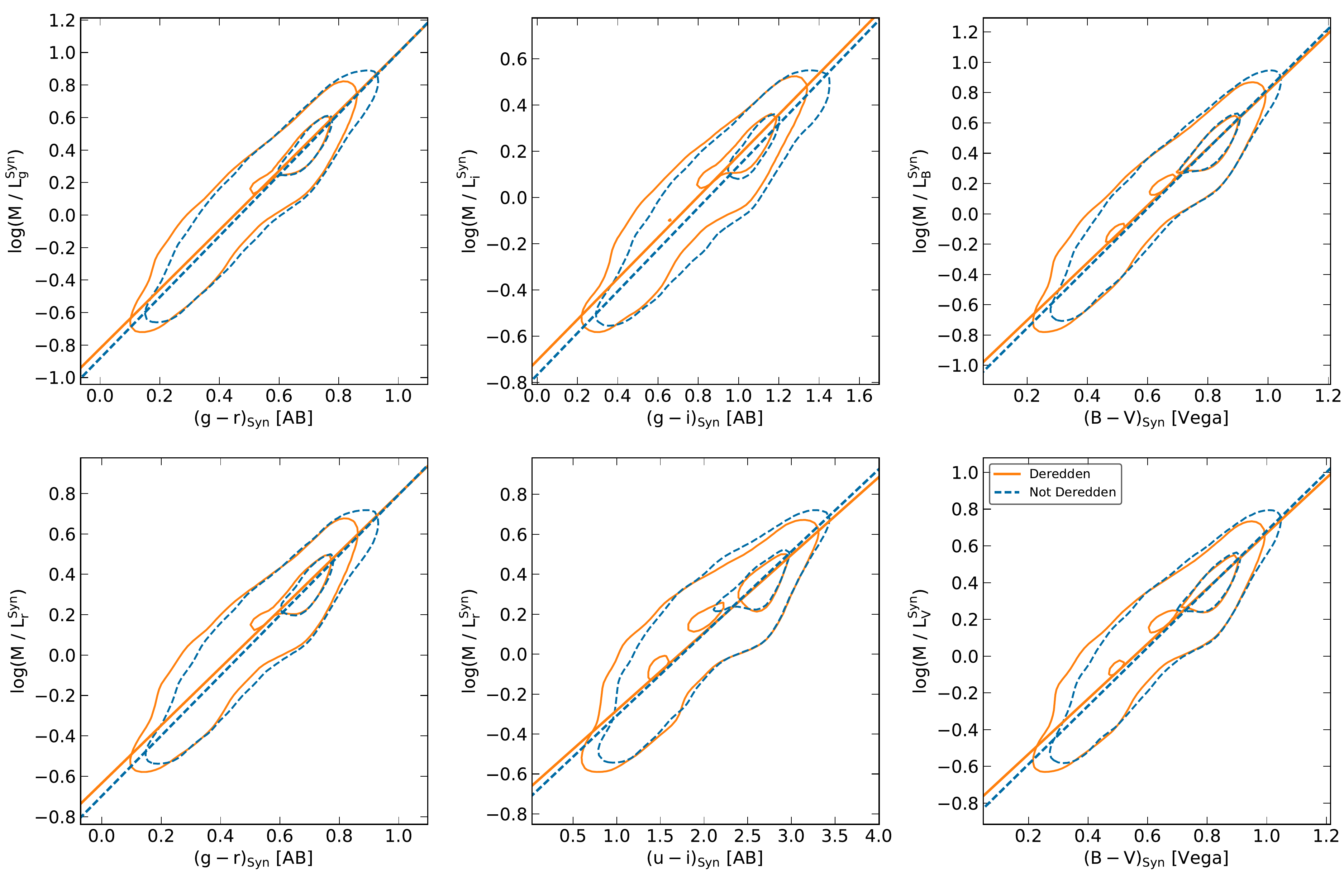}}
\caption{Relation between restframe color and \ml\ for different bands for not dereddened and dereddened
distributions. The contours represent the density distribution encompassing  90\% and 20\% of the points.}
\label{fig:MLCR_AV}
\end{figure*}

Figure \ref{fig:MLCR_AV} shows the \lmlc\ plane for some representative filter combinations. The 
contours follow the density distribution of the whole sample (encompassing 20\% and 90\% 
of the points) both for the not\textit{} dereddened (observed, the one provided in the tables) 
and dereddened colors and luminosities. The best linear fits for both distributions are also shown. 
As can be appreciated, the MLCRs do not change in a significant manner with extinction. The 
contours cover a similar area and location in the plane. A clear effect of the extinction is to 
distribute points along the linear relation, as seen by the 20\% contours.

\subsection{Effect of emission lines}

We briefly discussed in Sect. \ref{sec:mlradial} the mild effect of the emission lines. This effect can be 
studied by comparing the results from the synthetic and observed fits in the common bands available. 
The median differences in \ml\ in Sloan $g$ and $r$ bands, and $B$ and $V$ Johnson bands across the whole 
radial profile are on the order of 0.01 dex, generally below that value. The maximum difference is found 
for $r$ band for Sbc and Sd galaxies, with a median difference of 0.02 dex. The slightly larger difference 
is understood by the presence of H$\alpha$ emission in late-type galaxies. 

\section{Discussion}
\label{Sec:discussion}

The central goal of this paper is to provide an easy recipe to derive masses. In this 
final part we compare our results with previous relations found in the literature. We recall that all 
distributions and MLCRs (see section \ref{Sec:MLCRs}) in this section are computed in restframe in the 
not dereddened \textit{\textup{synthetic}} spectra.

\subsection{Comparison with previous MLCRs}

\begin{figure*}
\centering
\resizebox{\hsize}{!}{\includegraphics{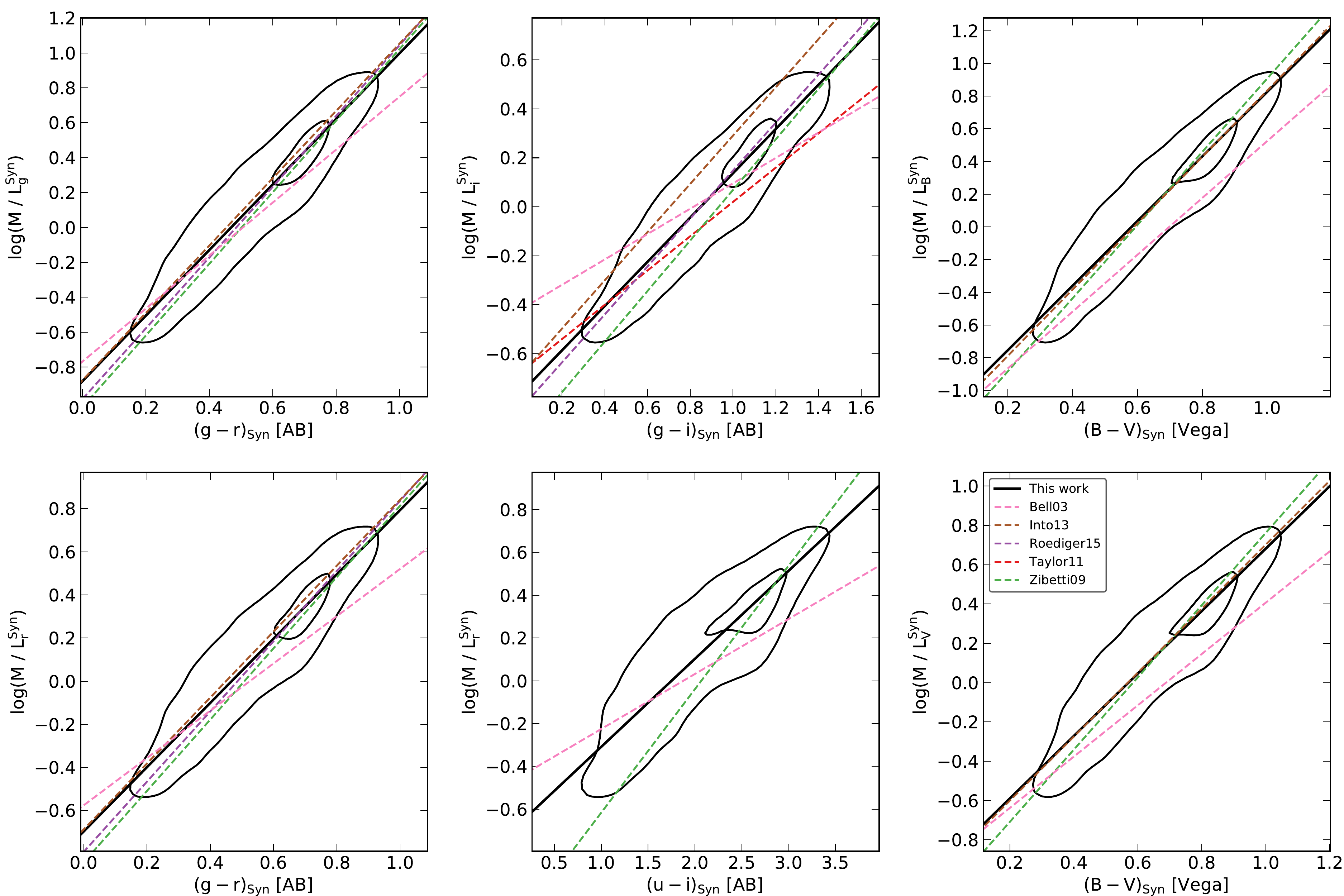}}
\caption{Comparison of the relation between restframe color and \ml\ for different 
bands for all galaxies to relations from the literature. The contours represent the 
density distribution encompassing  90\% and 20\% of the points.}
\label{fig:MLCR_av}
\end{figure*}

In Fig. \ref{fig:MLCR_av} we compare a few examples of our empirically calibrated \cml\ relation 
for different representative bands to other works found in the literature. Here we used the whole sample, 
including all types of galaxies. The (black) contours represent the density distribution encompassing  
90\% and 20\% of the points. The values of the slope and intercept associated with these plots can be found 
in Tables \ref{tb:AVTABLE_SynR_AB_px1} and \ref{tb:AVTABLE_SynR_Mag_px1}. Our spatially resolved MLCRs 
fits are plotted in solid black lines. 

Linear \cml\ relations have been computed by numerous studies in the past. One of the first one-colour 
methods was developed by \citet{Bell:2001} and later revised by \citet[][hereafter B03]{Bell:2003}. Their 
models, dust-free single exponential SFH libraries, are based on \citetalias{Bruzual:2003} SPS models. Thus,
 they do not account explicitly for dust and consider relatively smooth SFHs. In order to 
compare with our results, we have reduced the zero-points given in appendix A2 of
\citetalias{Bell:2003} by \oimf\ dex to transform from a ``diet-Salpeter IMF" used in their models, 
to a Chabrier IMF \citep[][and Sect. \ref{Sec:mstar} of this work]{GonzalezDelgado:2014a}. 
As has been reported by later works, the \citetalias{Bell:2003} relations 
present large discrepancies. Particularly, they strongly deviate toward lower values of \ml, with 
differences of a few dex in some filters (e.g. $g-i$). They tend to reproduce 
old unextincted stellar populations better (see below Fig.\ \ref{fig:MLCR_hub3}).

We also plot other common relations found in the literature. For example, the \cite{Into:2013} 
disk galaxy models (their Tables 4 and 5) run very close to ours, except for $g-i$, 
which overestimates \ml\ by $\sim$ 0.15 dex as compared to our results. We have adjusted 
their zero-points by applying a -0.025 dex offset to account for their 
Kroupa IMF \citep{Salim:2007}. 

Other relations close to our values are those of \citet{Roediger:2015} and \citet{Zibetti:2009}, 
although the latter has some discrepancies in the \mlbc{i}{g}{i} and particularly in the 
\mlbc{i}{g}{i} combination for low M/L ratios. Both works apply a least-squares regression to
the distributions followed by their model SFH libraries.

Finally, the \citet{Taylor:2011} relation, calibrated using SDSS $ugriz$ multi-band photometry of a 
large sample of galaxies, diverges from our relation in the \mlbc{i}{g}{i} plane for medium to 
high M/L values, that is, intermediate- and early-type galaxies, as we show in the next section. 
Their tight relation might be explained by the relatively young $\langle age \rangle_{mass}$
of their sample \citep{Zhang:2017}. 

These plots clearly show that the combination \mlbc{B}{B}{V} with Johnson-Cousin broadband filters is one of the best choices, both in terms of tightness of the distribution (scatter) 
and where the differences among the relations from the literature presented here have minimal 
differences. We explore the scatter in our relations in the next section.

In summary, our results are in agreement with previous results based on integrated M/L and colors of 
galaxies, with \mlb{g} and \mlb{r} being remarkably similar to the results from \citet{Zibetti:2009},
\citet{Into:2013}, and \citet{Roediger:2015}. The plane \mlbc{i}{g}{i} has more dispersion, 
but our results are similar to \citet{Roediger:2015}, and they are in between \citet{Zibetti:2009} 
and \citet{Into:2013} for ($g$-$i$) < 1, and \citet{Taylor:2011} and \citet{Into:2013} for redder 
colors. In the plane \mlbc{r}{u}{i}, our relation is in between  \citetalias{Bell:2003}  and 
\citet{Zibetti:2009}, which are, to our knowledge, the only two results from the literature in 
this filter combination. In the \mlbc{V}{B}{V} plane, our results are in perfect agreement with 
\citet{Roediger:2015}, and very close to \citet{Zibetti:2009}. As mentioned above, the relation 
\mlbc{B}{B}{V} is very tight, and all the results, ours included, agree very well, with 
the exception of \citetalias{Bell:2003}.

\subsection{Morphological MLCRs}

\begin{figure*}
\centering
\resizebox{\hsize}{!}{\includegraphics{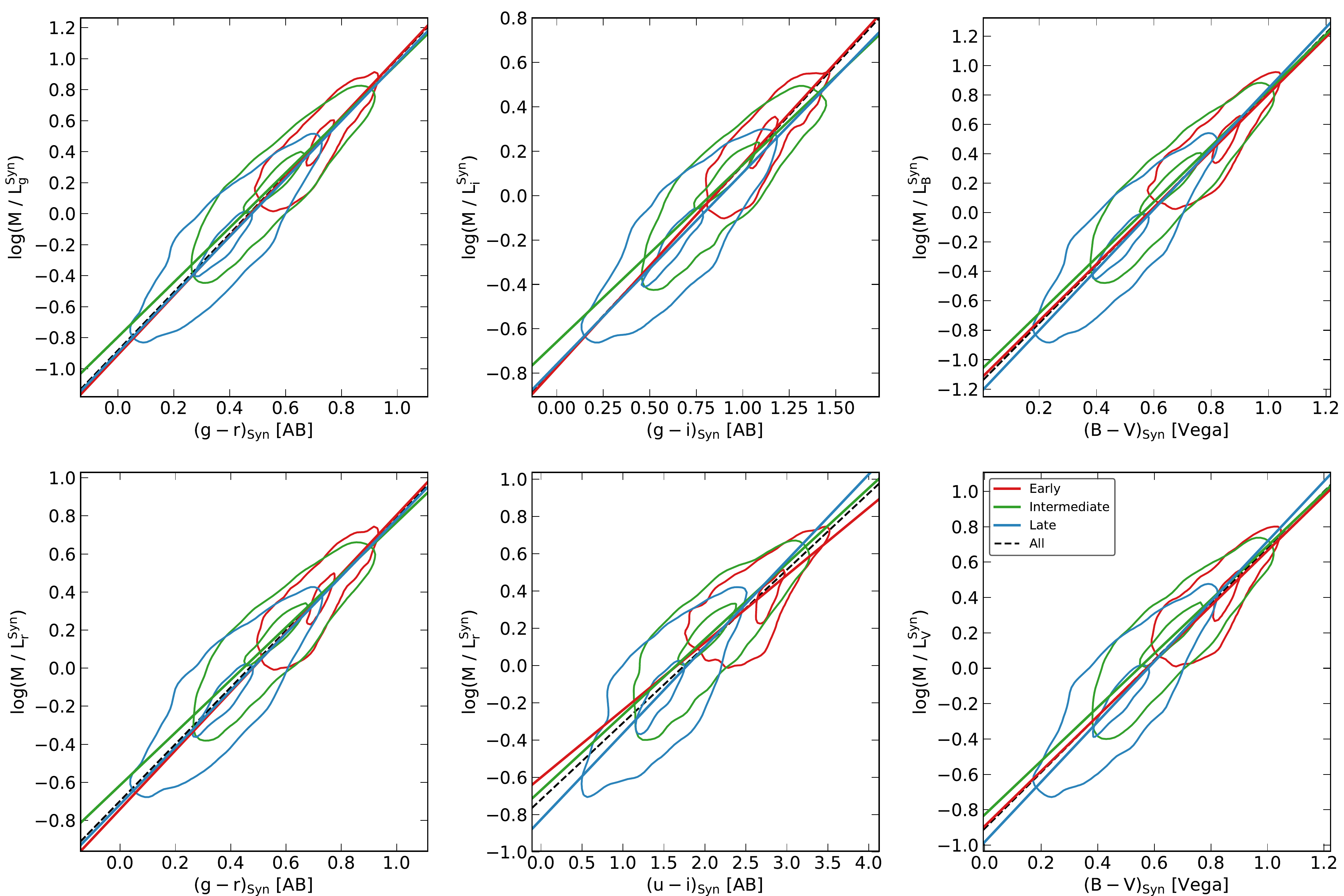}}
\caption{Relation between restframe color and \ml\ for different bands and for early (red), 
intermediate (green), and late galaxies (blue). Contours represent the density distribution 
encompassing  90\% and 20\% of the points. The dotted black line shows the best fit for all 
galaxy types.}
\label{fig:MLCR_hub3}
\end{figure*}

We now turn to the effect of the SFH on the \lmlc\ global relations. Different types of galaxies evolve in 
different ways, and thus, their locations in the \lmlc\ plane are different. In principle, one would 
expect the relations to be different for different types of galaxies. Figure \ref{fig:MLCR_hub3} 
compares the MLCRs at different bands for early- (red), intermediate- (green), 
and late-type galaxies (blue). The contours represent the density distribution encompassing 90\% and 20\% 
of the points. The dotted black line shows the best fit for all galaxy types (as in Fig. \ref{fig:MLCR_av}). 
The distribution of the three groups follows a sequential linear relation in the \lmlc\ plane. Early-type 
galaxies are located in the high \ml\ and red part of the diagram, with a fairly packed and tight 
distribution in most filters. Intermediate galaxies spread over a larger region, centered on an 
intermediate region in the \lmlc\ plane, but with a tail distribution that also encompasses similar values 
to those of early-type galaxies. Late-type galaxies display the widest and largest distribution of the 
three groups. They reach from very low values of \ml\ and blue colors up to relatively red colors and high 
\ml\ values.

The independent MLCRs for the three different groups are very close, in their range of validity, 
to the global MLCR for the whole sample. The differences between the different fits are larger 
when extrapolated outside the range of each distribution; thus, fits should be used only in 
their range of applicability. However, these differences are much smaller than the dispersion 
for each group distribution.

\subsection{Effect of the IMF}

The selection of SSPs can have an effect on the M/L as well. In principle, the main contributor 
to this difference would be an offset due to the IMF selection, but other secondary variables 
can also affect the slope and/or offset of the MLCRs, such as the range of metallicities 
introduced in the base or the stellar evolutionary tracks used for the SSPs. To check these 
possible variations, we also analyzed our dataset with the base we used in 
\citet[][hereafter {\rm GB}]{Garcia-Benito:2017}, which assumes a Salpeter IMF. 
In short, it consists of a combination of 254 SSPs. It combines the {\sc granada} models 
of \citet{GonzalezDelgado:2005} for ages younger than 60 Myr and the SSPs from 
\citet{Vazdekis:2015} based on BaSTi isochrones for older ages. The Z range covers eight 
metallicities, $\log Z/Z_{\odot} = -2.28$, $-1.79$, $-1.26$, $-0.66$, $-0.35$, $-0.06$, $0.25$, 
and $+0.40$.  The age is sampled by 37 SSPs per metallicity covering from 1 Myr to 14 Gyr. The 
IMF is the Salpeter IMF. Dust effects were modeled in the same way as our fiducial 
results: as a foreground screen with a \citet{Cardelli:1989} reddening law with R$_V$ = 3.1.

\begin{figure*}
\centering
\resizebox{\hsize}{!}{\includegraphics{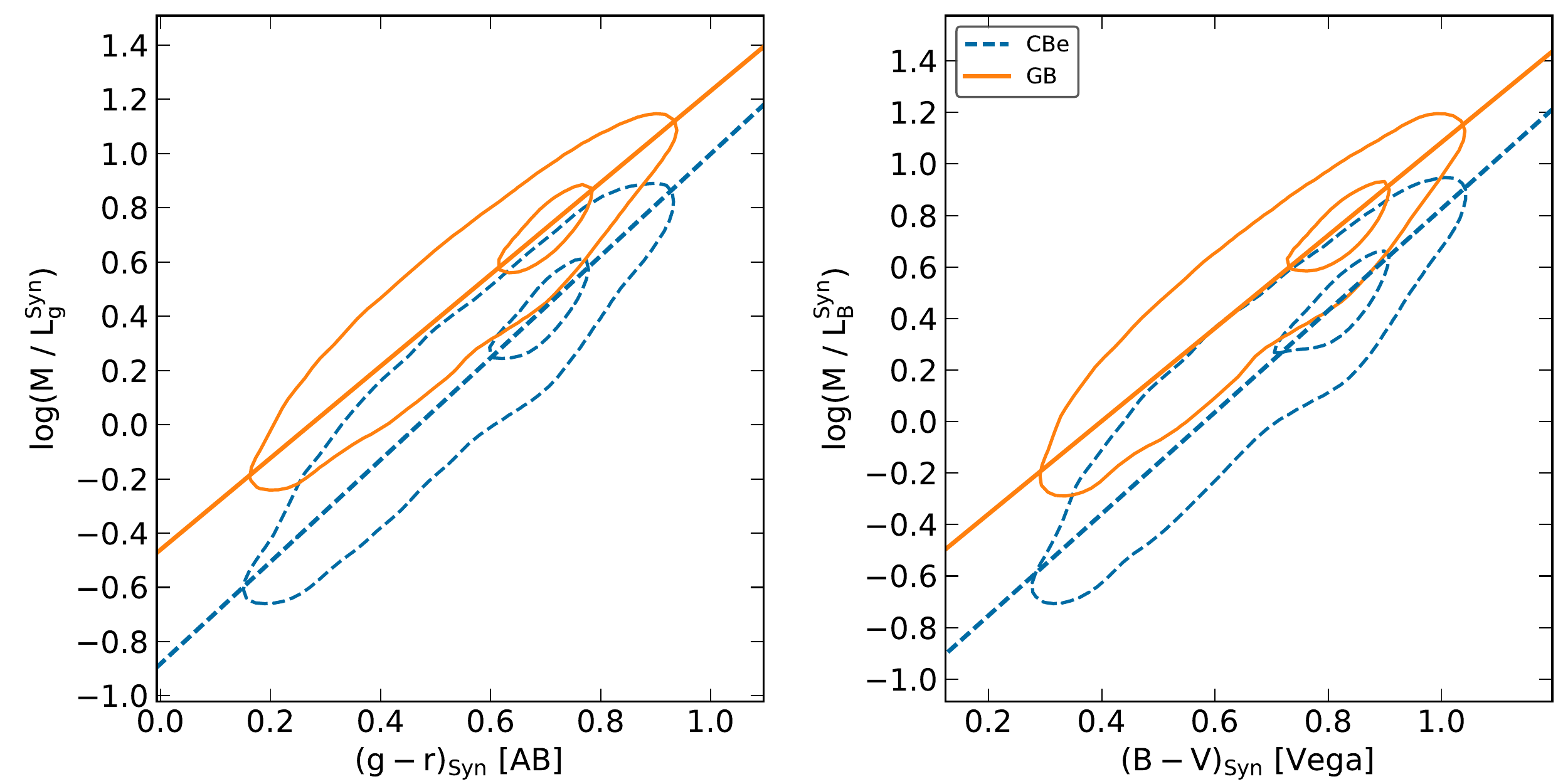}}
\caption{Comparison of the relation between restframe color and \ml\ for all galaxies 
for bases {\rm CBe} (Chabrier IMF) and {\rm GB} (Salpeter IMF). Contours represent 
the density distribution encompassing 90\% and 20\% of the points.}
\label{fig:MLCR_bases}
\end{figure*}

Figure \ref{fig:MLCR_bases} compares the MLCRs obtained using both sets of SSPs including the 
whole sample in the derivations of the relations. The contours represent the density distribution 
encompassing 90\% and 20\% of the points. The wavelength range of base {\rm GB} is 
shorter than {\rm CBe}, covering from 3500 \AA\ to 7000 \AA. Thus, we only show two examples of 
the common filters available in both results. As expected, the distribution using {\rm GB} is 
located higher in the \lmlc\ plane due to the choice of a Salpeter IMF. However, in addition to the 
offset in the zero-points, there are some differences for very extreme colors due to the 
differences in the slopes. For blue colors, the relations can give M/L differences as high as 
0.4 dex, while for red colors (high M/L values), this difference is on the order of 0.2 dex. 
Both samples have nearly the same scatter (i.e., uncertainties in the MLCRs) in the distributions.

\section{Summary and conclusions}
\label{Sec:Summary}

We have applied the fossil record method of the stellar population to a sample of \ngal\ galaxies observed 
with integral field spectroscopy in the CALIFA survey to derive the radial structure of the M/L ratio. 
Observed and synthetic radial colors in  optical broadbands were also measured to study the 
spatially resolved MLCRs. Our sample covers a wide range of morphological types that includes early -type galaxies (E, S0) and spirals (Sa, Sb, Sc, Sd), and galaxy stellar masses ranges from 
10$^{8.4}$ to 10$^{12}$ M$_\odot$. Our main results are listed below.

\begin{enumerate}

\item The radial profile of $M/L$ scales with Hubble type, decreasing outward in all cases. The 
inner gradient of M/L (within 1 HLR) is steeper than the outer gradient. The gradient is steeper in 
Sb-Sbc than in early types (E, S0, and Sa); late-type spirals (Sc-Sd) show the flattest gradient. 
These trends are independent of the photometric band used for the luminosity.
 
\item The spatially resolved structure of the $M/L$ ratio and colors up to 2 HLR was used to derive the 
MLCRs. A linear relation was derived for several pairs of $M/L$-optical colors. The mean scatter of all 
the combinations of filters is $\sim 0.13$ dex. A scatter lower that 0.1 dex is obtained  for pairs such 
as ($i$, $g-r$) and ($R$, $B-R$). 
 
\item Uncertainties associated with the effect of emission lines in the luminosity band and optical colors 
are very small, with the maximum difference in M/L of $\sim$ 0.02 dex, found in the \textit{r}  $    band$ and for 
intermediate- and late-type galaxies, from which the largest contribution of the H$\alpha$ emission 
is expected in $L_r$. The effect of extinction is also negligible because, as pointed  out 
by \citetalias{Bell:2003}, the reddening correction vector runs parallel to the  MLCRs at optical 
wavelengths.
 
\item The MLCRs are derived for galaxies of different Hubble types. These relations are very close, 
in their range of validity, to the relations found for the whole sample. The largest difference 
occurs for the pair ($r$, $u-i$) with an offset of 0.6 for late-type galaxies ($u-i$) < 0.5 and 
0.05 for early-type galaxies with ($u-i$) > 3.5.

\item The effect of the SSPs used for the full spectral fitting with  {\stl} was also studied. 
We compared our results in $M_\star$ and MLCRs derived with the bases {\rm CBe} and {\rm GB}. 
In addition to the differences in the evolutionary codes (\citetalias{Bruzual:2003} vs. \citealt{Vazdekis:2015}) 
and the corresponding stellar evolution, these two sets of SSPs have different IMFs 
(Chabrier vs. Salpeter). The corresponding MLCRs mainly reflect the difference in IMF, with a shift 
to larger $M/L$ of $\sim$ \oimf\ for SSPs with Salpeter IMF. However, differences in the SFH give a 
random error of $\geq$ 0.4 dex for galaxies bluer than ($g-r$) = 0.2 and $\leq$ 0.2 dex for galaxies 
redder than ($g-r$) = 0.9.
 
\item The comparison of the MLCRs with other published results indicates that our relations 
are compatible with results based on integrated $M/L$ and color of galaxies. In particular, 
the relation \mlbc{B}{B}{V} is very tight and perfectly agrees with them, with the exception 
of \citetalias{Bell:2003}. The relations of $M/L_g$ and $M/L_r$ 
with ($g-r$) are remarkably similar to the relations from \citet{Zibetti:2009}, \citet{Into:2013}, 
and \citet{Roediger:2015}. However, the \mlbc{i}{g}{i} plane shows greater dispersion between 
all the results in the literature, but our relations are similar to those of \citet{Roediger:2015}, and 
they are in between those of \citet{Zibetti:2009} and \citet{Into:2013} for ($g$-$i$) < 1, and 
\citet{Taylor:2011} and \citet{Into:2013} for redder colors.

\end{enumerate}

\begin{acknowledgements}
CALIFA is the first legacy survey carried out at Calar Alto. The CALIFA collaboration would 
like to thank the IAA-CSIC and MPIA-MPG as major partners of the observatory, and CAHA itself, 
for the unique access to telescope time and support in manpower and infrastructures. We also 
thank the CAHA staff for the dedication to this project. We thank the support of the IAA 
Computing group. 
Support from the Spanish Ministerio de Econom\'ia y Competitividad, through projects 
AYA2016-77846-P, AYA2014-57490-P, AYA2010-15081, and Junta de Andaluc\'ia P12-FQM-2828. 
SFS is grateful for the support of a CONACYT (Mexico) grant CB-285080, and funding 
from the PAPIIT-DGAPA-IA101217 (UNAM).
This research made use of Python 
(\href{http://www.python.org}{http://www.python.org}); Numpy \citep{vanDerWalt:2011}, 
Astropy \citep{Astropy:2013}, Pandas \citep{McKinney:2011}, Matplotlib \citep{Hunter:2007}, 
and Seaborn \citep{Waskom:2016}. We thank the referee for very useful comments that
improved the presentation of the paper.
\end{acknowledgements}

\bibliographystyle{aa}
\bibliography{rgb_califa_ML}

\begin{table*}
\caption{Linear fitting at 0 $\leq$ R$_{HLR}$ $\leq$ 2 to $log(M/L_{\lambda}) = a_{\lambda} + (b_{\lambda} \times color)$ measured in restframe in the synthetic spectra (luminosities \textit{not} corrected for reddening).}
\label{tb:AVTABLE_SynR_AB_px1}
\centering
\begin{tabular}{lllllllllllllllll}
\hline\hline
   Color & $a_{u}$ & $b_{u}$ & $\sigma_{u}$ & $a_{g}$ & $b_{g}$ & $\sigma_{g}$ & $a_{r}$ & $b_{r}$ & $\sigma_{r}$ & $a_{i}$ & $b_{i}$ & $\sigma_{i}$ & $a_{z}$ & $b_{z}$ & $\sigma_{z}$ \\
\hline

\multicolumn{16}{c}{All galaxies}\\\hline
 $u - g$ &   -1.19 &    1.14 &         0.19 &   -0.73 &    0.77 &         0.19 &   -0.57 &    0.60 &         0.16 &   -0.58 &    0.54 &         0.14 &   -0.63 &    0.49 &         0.13 \\
 $u - r$ &   -1.30 &    0.83 &         0.14 &   -0.83 &    0.58 &         0.15 &   -0.65 &    0.45 &         0.13 &   -0.65 &    0.40 &         0.12 &   -0.69 &    0.37 &         0.11 \\
 $u - i$ &   -1.40 &    0.74 &         0.12 &   -0.91 &    0.52 &         0.14 &   -0.72 &    0.41 &         0.12 &   -0.71 &    0.36 &         0.11 &   -0.74 &    0.33 &         0.11 \\
 $u - z$ &   -1.48 &    0.69 &         0.11 &   -0.97 &    0.48 &         0.13 &   -0.77 &    0.38 &         0.12 &   -0.75 &    0.34 &         0.11 &   -0.78 &    0.31 &         0.10 \\
 $g - r$ &   -1.30 &    2.59 &         0.11 &   -0.88 &    1.88 &         0.10 &   -0.70 &    1.49 &         0.10 &   -0.69 &    1.31 &         0.09 &   -0.72 &    1.19 &         0.09 \\
 $g - i$ &   -1.44 &    1.77 &         0.13 &   -0.99 &    1.29 &         0.10 &   -0.79 &    1.03 &         0.09 &   -0.77 &    0.90 &         0.09 &   -0.79 &    0.82 &         0.09 \\
 $g - z$ &   -1.52 &    1.43 &         0.15 &   -1.06 &    1.05 &         0.11 &   -0.84 &    0.83 &         0.10 &   -0.81 &    0.73 &         0.10 &   -0.83 &    0.66 &         0.10 \\
 $r - i$ &   -1.53 &    5.01 &         0.22 &   -1.08 &    3.74 &         0.15 &   -0.86 &    2.98 &         0.13 &   -0.83 &    2.60 &         0.12 &   -0.84 &    2.34 &         0.12 \\
 $r - z$ &   -1.58 &    2.86 &         0.24 &   -1.12 &    2.13 &         0.16 &   -0.89 &    1.70 &         0.14 &   -0.85 &    1.48 &         0.13 &   -0.86 &    1.32 &         0.13 \\
 $i - z$ &   -1.49 &    6.07 &         0.29 &   -1.05 &    4.52 &         0.20 &   -0.83 &    3.59 &         0.17 &   -0.80 &    3.11 &         0.16 &   -0.80 &    2.76 &         0.15 \\
\hline
\end{tabular}
\tablefoot{Linear fitting at 0 $\leq$ R$_{HLR}$ $\leq$ 2 to optical log(M/L$_{\lambda}$)-color with base CBe and Chabrier IMF. The monochromatic mass-to-light ratios $(M/L_{\lambda})$ are in solar units. The SDSS $ugriz$ filters are in the AB magnitude system. $\sigma_{\lambda}$ is the scatter of the residuals of the relation log(M/L$_{\lambda}$)-color.}
\end{table*}
\begin{table*}
\caption{Linear fitting at 0 $\leq$ R$_{HLR}$ $\leq$ 2 to $log(M/L_{\lambda}) = a_{\lambda} + (b_{\lambda} \times color)$ measured in restframe in the synthetic spectra (luminosities \textit{not} corrected for reddening).}
\label{tb:AVTABLE_SynR_Mag_px1}
\centering
\begin{tabular}{lllllllllll}
\hline\hline
   Color & $a_{B}$ & $b_{B}$ & $\sigma_{B}$ & $a_{V}$ & $b_{V}$ & $\sigma_{V}$ & $a_{R}$ & $b_{R}$ & $\sigma_{R}$ \\
\hline

\multicolumn{10}{c}{All galaxies}\\\hline
 $B - V$ &   -1.15 &    1.97 &         0.11 &   -0.91 &    1.59 &         0.11 &   -0.82 &    1.36 &         0.10 \\
 $B - R$ &   -1.45 &    1.25 &         0.10 &   -1.15 &    1.01 &         0.10 &   -1.03 &    0.86 &         0.09 \\
 $V - R$ &   -1.83 &    3.22 &         0.12 &   -1.47 &    2.61 &         0.11 &   -1.30 &    2.23 &         0.10 \\
\hline
\end{tabular}
\tablefoot{Linear fitting at 0 $\leq$ R$_{HLR}$ $\leq$ 2 to optical log(M/L$_{\lambda}$)-color with base CBe and Chabrier IMF. The monochromatic mass-to-light ratios $(M/L_{\lambda})$ are in solar units. The Johnson-Cousins $BVR$ filters are in the Vega magnitude system. $\sigma_{\lambda}$ is the scatter of the residuals of the relation log(M/L$_{\lambda}$)-color.}
\end{table*}
\begin{table*}
\caption{Linear fitting at 0 $\leq$ R$_{HLR}$ $\leq$ 2 to $log(M/L_{\lambda}) = a_{\lambda} + (b_{\lambda} \times color)$ measured in restframe in the synthetic spectra (luminosities \textit{not} corrected for reddening).}
\label{tb:hub3_SynR_AB_px1}
\centering
\begin{tabular}{lllllllllllllllll}
\hline\hline
   Color & $a_{u}$ & $b_{u}$ & $\sigma_{u}$ & $a_{g}$ & $b_{g}$ & $\sigma_{g}$ & $a_{r}$ & $b_{r}$ & $\sigma_{r}$ & $a_{i}$ & $b_{i}$ & $\sigma_{i}$ & $a_{z}$ & $b_{z}$ & $\sigma_{z}$ \\
\hline

\multicolumn{16}{c}{Early galaxies}\\\hline
 $u - g$ &   -0.81 &    0.89 &         0.16 &   -0.33 &    0.51 &         0.16 &   -0.25 &    0.40 &         0.14 &   -0.31 &    0.36 &         0.12 &   -0.38 &    0.33 &         0.11 \\
 $u - r$ &   -1.13 &    0.75 &         0.13 &   -0.59 &    0.46 &         0.14 &   -0.46 &    0.37 &         0.12 &   -0.49 &    0.33 &         0.11 &   -0.54 &    0.30 &         0.10 \\
 $u - i$ &   -1.32 &    0.71 &         0.11 &   -0.76 &    0.46 &         0.12 &   -0.60 &    0.36 &         0.11 &   -0.61 &    0.32 &         0.10 &   -0.64 &    0.29 &         0.09 \\
 $u - z$ &   -1.40 &    0.66 &         0.10 &   -0.85 &    0.44 &         0.12 &   -0.68 &    0.35 &         0.11 &   -0.67 &    0.31 &         0.10 &   -0.68 &    0.27 &         0.09 \\
 $g - r$ &   -1.35 &    2.70 &         0.10 &   -0.91 &    1.92 &         0.09 &   -0.74 &    1.55 &         0.09 &   -0.72 &    1.35 &         0.08 &   -0.74 &    1.21 &         0.08 \\
 $g - i$ &   -1.40 &    1.78 &         0.12 &   -0.98 &    1.29 &         0.09 &   -0.81 &    1.05 &         0.08 &   -0.77 &    0.92 &         0.08 &   -0.77 &    0.81 &         0.08 \\
 $g - z$ &   -1.38 &    1.37 &         0.13 &   -0.97 &    1.00 &         0.09 &   -0.80 &    0.82 &         0.08 &   -0.76 &    0.71 &         0.08 &   -0.76 &    0.62 &         0.08 \\
 $r - i$ &   -1.03 &    4.04 &         0.18 &   -0.79 &    3.14 &         0.12 &   -0.67 &    2.58 &         0.10 &   -0.64 &    2.23 &         0.10 &   -0.65 &    1.95 &         0.10 \\
 $r - z$ &   -1.02 &    2.24 &         0.19 &   -0.77 &    1.73 &         0.12 &   -0.65 &    1.42 &         0.11 &   -0.62 &    1.22 &         0.10 &   -0.62 &    1.05 &         0.10 \\
 $i - z$ &   -0.77 &    4.32 &         0.21 &   -0.58 &    3.32 &         0.14 &   -0.49 &    2.71 &         0.12 &   -0.48 &    2.30 &         0.11 &   -0.48 &    1.96 &         0.11 \\
\hline
\multicolumn{16}{c}{Intermediate galaxies}\\\hline
 $u - g$ &   -1.13 &    1.14 &         0.17 &   -0.67 &    0.77 &         0.17 &   -0.51 &    0.59 &         0.15 &   -0.52 &    0.52 &         0.13 &   -0.56 &    0.46 &         0.12 \\
 $u - r$ &   -1.26 &    0.83 &         0.12 &   -0.80 &    0.58 &         0.14 &   -0.61 &    0.45 &         0.12 &   -0.60 &    0.39 &         0.11 &   -0.63 &    0.35 &         0.10 \\
 $u - i$ &   -1.35 &    0.73 &         0.11 &   -0.87 &    0.52 &         0.12 &   -0.67 &    0.41 &         0.11 &   -0.65 &    0.35 &         0.10 &   -0.68 &    0.31 &         0.10 \\
 $u - z$ &   -1.40 &    0.66 &         0.10 &   -0.92 &    0.48 &         0.11 &   -0.71 &    0.37 &         0.10 &   -0.68 &    0.32 &         0.10 &   -0.70 &    0.28 &         0.10 \\
 $g - r$ &   -1.17 &    2.36 &         0.10 &   -0.80 &    1.76 &         0.09 &   -0.62 &    1.38 &         0.09 &   -0.61 &    1.19 &         0.08 &   -0.63 &    1.06 &         0.08 \\
 $g - i$ &   -1.26 &    1.56 &         0.11 &   -0.87 &    1.18 &         0.09 &   -0.68 &    0.92 &         0.09 &   -0.66 &    0.80 &         0.09 &   -0.68 &    0.70 &         0.09 \\
 $g - z$ &   -1.30 &    1.22 &         0.13 &   -0.91 &    0.93 &         0.10 &   -0.71 &    0.73 &         0.09 &   -0.68 &    0.63 &         0.09 &   -0.69 &    0.55 &         0.09 \\
 $r - i$ &   -1.24 &    4.04 &         0.18 &   -0.88 &    3.13 &         0.13 &   -0.69 &    2.47 &         0.11 &   -0.66 &    2.11 &         0.11 &   -0.67 &    1.83 &         0.11 \\
 $r - z$ &   -1.25 &    2.24 &         0.19 &   -0.89 &    1.74 &         0.14 &   -0.69 &    1.37 &         0.12 &   -0.66 &    1.16 &         0.11 &   -0.66 &    1.00 &         0.11 \\
 $i - z$ &   -1.11 &    4.54 &         0.22 &   -0.79 &    3.53 &         0.17 &   -0.61 &    2.76 &         0.14 &   -0.58 &    2.33 &         0.13 &   -0.59 &    1.97 &         0.13 \\
\hline
\multicolumn{16}{c}{Late galaxies}\\\hline
 $u - g$ &   -1.27 &    1.13 &         0.21 &   -0.84 &    0.81 &         0.21 &   -0.66 &    0.63 &         0.18 &   -0.65 &    0.54 &         0.16 &   -0.69 &    0.49 &         0.15 \\
 $u - r$ &   -1.38 &    0.86 &         0.16 &   -0.96 &    0.65 &         0.17 &   -0.76 &    0.51 &         0.15 &   -0.74 &    0.44 &         0.14 &   -0.77 &    0.40 &         0.13 \\
 $u - i$ &   -1.46 &    0.76 &         0.14 &   -1.04 &    0.59 &         0.15 &   -0.83 &    0.46 &         0.14 &   -0.80 &    0.40 &         0.13 &   -0.81 &    0.36 &         0.13 \\
 $u - z$ &   -1.52 &    0.70 &         0.13 &   -1.09 &    0.54 &         0.15 &   -0.87 &    0.43 &         0.14 &   -0.83 &    0.37 &         0.13 &   -0.84 &    0.33 &         0.12 \\
 $g - r$ &   -1.21 &    2.31 &         0.12 &   -0.90 &    1.87 &         0.12 &   -0.72 &    1.50 &         0.11 &   -0.70 &    1.30 &         0.11 &   -0.73 &    1.16 &         0.11 \\
 $g - i$ &   -1.31 &    1.52 &         0.13 &   -0.98 &    1.25 &         0.12 &   -0.79 &    1.00 &         0.11 &   -0.76 &    0.86 &         0.11 &   -0.78 &    0.77 &         0.11 \\
 $g - z$ &   -1.37 &    1.20 &         0.14 &   -1.03 &    0.99 &         0.13 &   -0.83 &    0.79 &         0.12 &   -0.79 &    0.68 &         0.11 &   -0.81 &    0.60 &         0.11 \\
 $r - i$ &   -1.32 &    3.82 &         0.20 &   -1.01 &    3.20 &         0.16 &   -0.81 &    2.57 &         0.14 &   -0.78 &    2.20 &         0.13 &   -0.79 &    1.95 &         0.13 \\
 $r - z$ &   -1.37 &    2.16 &         0.21 &   -1.04 &    1.79 &         0.17 &   -0.83 &    1.44 &         0.15 &   -0.79 &    1.23 &         0.14 &   -0.80 &    1.08 &         0.14 \\
 $i - z$ &   -1.27 &    4.30 &         0.24 &   -0.96 &    3.54 &         0.20 &   -0.76 &    2.83 &         0.17 &   -0.73 &    2.39 &         0.16 &   -0.74 &    2.07 &         0.15 \\
\hline
\end{tabular}
\tablefoot{Linear fitting at 0 $\leq$ R$_{HLR}$ $\leq$ 2 to optical log(M/L$_{\lambda}$)-color with base CBe and Chabrier IMF. The monochromatic mass-to-light ratios $(M/L_{\lambda})$ are in solar units. The SDSS $ugriz$ filters are in the AB magnitude system. $\sigma_{\lambda}$ is the scatter of the residuals of the relation log(M/L$_{\lambda}$)-color.}
\end{table*}
\begin{table*}
\caption{Linear fitting at 0 $\leq$ R$_{HLR}$ $\leq$ 2 to $log(M/L_{\lambda}) = a_{\lambda} + (b_{\lambda} \times color)$ measured in restframe in the synthetic spectra (luminosities \textit{not} corrected for reddening).}
\label{tb:hub3_SynR_Mag_px1}
\centering
\begin{tabular}{lllllllllll}
\hline\hline
   Color & $a_{B}$ & $b_{B}$ & $\sigma_{B}$ & $a_{V}$ & $b_{V}$ & $\sigma_{V}$ & $a_{R}$ & $b_{R}$ & $\sigma_{R}$ \\
\hline

\multicolumn{10}{c}{Early galaxies}\\\hline
 $B - V$ &   -1.12 &    1.93 &         0.10 &   -0.89 &    1.56 &         0.10 &   -0.81 &    1.33 &         0.09 \\
 $B - R$ &   -1.46 &    1.26 &         0.09 &   -1.19 &    1.03 &         0.08 &   -1.06 &    0.88 &         0.08 \\
 $V - R$ &   -1.62 &    2.96 &         0.11 &   -1.34 &    2.46 &         0.09 &   -1.19 &    2.10 &         0.09 \\
\hline
\multicolumn{10}{c}{Intermediate galaxies}\\\hline
 $B - V$ &   -1.06 &    1.90 &         0.10 &   -0.83 &    1.52 &         0.10 &   -0.74 &    1.28 &         0.09 \\
 $B - R$ &   -1.32 &    1.17 &         0.09 &   -1.03 &    0.94 &         0.09 &   -0.91 &    0.79 &         0.08 \\
 $V - R$ &   -1.57 &    2.81 &         0.11 &   -1.25 &    2.27 &         0.10 &   -1.09 &    1.90 &         0.10 \\
\hline
\multicolumn{10}{c}{Late galaxies}\\\hline
 $B - V$ &   -1.21 &    2.06 &         0.13 &   -0.98 &    1.70 &         0.13 &   -0.88 &    1.43 &         0.12 \\
 $B - R$ &   -1.47 &    1.26 &         0.12 &   -1.20 &    1.04 &         0.12 &   -1.06 &    0.88 &         0.11 \\
 $V - R$ &   -1.71 &    2.92 &         0.13 &   -1.41 &    2.43 &         0.12 &   -1.23 &    2.05 &         0.12 \\
\hline
\end{tabular}
\tablefoot{Linear fitting at 0 $\leq$ R$_{HLR}$ $\leq$ 2 to optical log(M/L$_{\lambda}$)-color with base CBe and Chabrier IMF. The monochromatic mass-to-light ratios $(M/L_{\lambda})$ are in solar units. The Johnson-Cousins $BVR$ filters are in the Vega magnitude system. $\sigma_{\lambda}$ is the scatter of the residuals of the relation log(M/L$_{\lambda}$)-color.}
\end{table*}
\longtab{
\begin{longtable}{llllllllllllllll}
\caption{\label{tb:hubtyp_SynR_AB_px1} Linear fitting at 0 $\leq$ R$_{HLR}$ $\leq$ 2 to $log(M/L_{\lambda}) = a_{\lambda} + (b_{\lambda} \times color)$ measured in restframe in the synthetic spectra (luminosities \textit{not} corrected for reddening).}\\
\hline\hline
   Color & $a_{u}$ & $b_{u}$ & $\sigma_{u}$ & $a_{g}$ & $b_{g}$ & $\sigma_{g}$ & $a_{r}$ & $b_{r}$ & $\sigma_{r}$ & $a_{i}$ & $b_{i}$ & $\sigma_{i}$ & $a_{z}$ & $b_{z}$ & $\sigma_{z}$ \\
\hline
\endfirsthead
\caption{continued.}\\
\hline\hline
   Color & $a_{u}$ & $b_{u}$ & $\sigma_{u}$ & $a_{g}$ & $b_{g}$ & $\sigma_{g}$ & $a_{r}$ & $b_{r}$ & $\sigma_{r}$ & $a_{i}$ & $b_{i}$ & $\sigma_{i}$ & $a_{z}$ & $b_{z}$ & $\sigma_{z}$ \\
\endhead
\hline
\endfoot
\multicolumn{16}{c}{E galaxies}\\\hline
 $u - g$ &   -1.71 &    1.40 &         0.14 &   -1.19 &    1.00 &         0.13 &   -1.02 &    0.83 &         0.12 &   -0.96 &    0.73 &         0.11 &   -0.93 &    0.64 &         0.11 \\
 $u - r$ &   -1.94 &    1.07 &         0.10 &   -1.43 &    0.79 &         0.11 &   -1.23 &    0.67 &         0.10 &   -1.15 &    0.58 &         0.09 &   -1.11 &    0.52 &         0.09 \\
 $u - i$ &   -1.95 &    0.92 &         0.09 &   -1.47 &    0.69 &         0.10 &   -1.27 &    0.58 &         0.09 &   -1.18 &    0.51 &         0.09 &   -1.14 &    0.46 &         0.08 \\
 $u - z$ &   -1.90 &    0.81 &         0.08 &   -1.43 &    0.61 &         0.09 &   -1.24 &    0.52 &         0.09 &   -1.16 &    0.46 &         0.08 &   -1.12 &    0.41 &         0.08 \\
 $g - r$ &   -1.81 &    3.33 &         0.08 &   -1.43 &    2.60 &         0.08 &   -1.26 &    2.22 &         0.07 &   -1.19 &    1.97 &         0.07 &   -1.17 &    1.78 &         0.07 \\
 $g - i$ &   -1.71 &    2.07 &         0.08 &   -1.36 &    1.62 &         0.07 &   -1.20 &    1.39 &         0.07 &   -1.13 &    1.23 &         0.06 &   -1.11 &    1.11 &         0.07 \\
 $g - z$ &   -1.63 &    1.57 &         0.08 &   -1.29 &    1.22 &         0.07 &   -1.14 &    1.05 &         0.07 &   -1.08 &    0.93 &         0.07 &   -1.06 &    0.83 &         0.07 \\
 $r - i$ &   -1.28 &    4.79 &         0.10 &   -1.04 &    3.79 &         0.08 &   -0.92 &    3.24 &         0.08 &   -0.89 &    2.86 &         0.07 &   -0.89 &    2.57 &         0.07 \\
 $r - z$ &   -1.26 &    2.66 &         0.11 &   -1.02 &    2.09 &         0.09 &   -0.90 &    1.79 &         0.08 &   -0.86 &    1.57 &         0.08 &   -0.86 &    1.40 &         0.08 \\
 $i - z$ &   -1.06 &    5.42 &         0.13 &   -0.85 &    4.23 &         0.11 &   -0.75 &    3.60 &         0.10 &   -0.72 &    3.14 &         0.09 &   -0.72 &    2.77 &         0.09 \\
\hline
\multicolumn{16}{c}{S0 galaxies}\\\hline
 $u - g$ &   -1.03 &    1.02 &         0.15 &   -0.55 &    0.63 &         0.15 &   -0.43 &    0.50 &         0.13 &   -0.45 &    0.44 &         0.12 &   -0.49 &    0.39 &         0.11 \\
 $u - r$ &   -1.32 &    0.82 &         0.12 &   -0.82 &    0.55 &         0.13 &   -0.66 &    0.44 &         0.12 &   -0.65 &    0.39 &         0.11 &   -0.66 &    0.34 &         0.10 \\
 $u - i$ &   -1.45 &    0.75 &         0.10 &   -0.95 &    0.52 &         0.12 &   -0.78 &    0.42 &         0.11 &   -0.75 &    0.37 &         0.10 &   -0.75 &    0.33 &         0.09 \\
 $u - z$ &   -1.50 &    0.69 &         0.10 &   -1.00 &    0.48 &         0.11 &   -0.83 &    0.39 &         0.10 &   -0.79 &    0.34 &         0.09 &   -0.79 &    0.30 &         0.09 \\
 $g - r$ &   -1.44 &    2.83 &         0.09 &   -1.01 &    2.05 &         0.09 &   -0.85 &    1.68 &         0.09 &   -0.82 &    1.49 &         0.08 &   -0.83 &    1.33 &         0.08 \\
 $g - i$ &   -1.49 &    1.87 &         0.09 &   -1.08 &    1.38 &         0.08 &   -0.91 &    1.14 &         0.08 &   -0.88 &    1.01 &         0.07 &   -0.87 &    0.90 &         0.07 \\
 $g - z$ &   -1.48 &    1.45 &         0.10 &   -1.08 &    1.08 &         0.08 &   -0.92 &    0.90 &         0.08 &   -0.88 &    0.79 &         0.07 &   -0.87 &    0.70 &         0.07 \\
 $r - i$ &   -1.18 &    4.48 &         0.14 &   -0.92 &    3.48 &         0.10 &   -0.79 &    2.91 &         0.09 &   -0.77 &    2.55 &         0.08 &   -0.77 &    2.28 &         0.08 \\
 $r - z$ &   -1.21 &    2.55 &         0.15 &   -0.92 &    1.95 &         0.10 &   -0.79 &    1.63 &         0.09 &   -0.76 &    1.42 &         0.09 &   -0.76 &    1.26 &         0.09 \\
 $i - z$ &   -1.03 &    5.21 &         0.17 &   -0.77 &    3.95 &         0.12 &   -0.66 &    3.28 &         0.11 &   -0.64 &    2.85 &         0.10 &   -0.64 &    2.49 &         0.10 \\
\hline
\multicolumn{16}{c}{Sa galaxies}\\\hline
 $u - g$ &   -0.73 &    0.88 &         0.16 &   -0.27 &    0.51 &         0.15 &   -0.20 &    0.40 &         0.13 &   -0.26 &    0.36 &         0.11 &   -0.34 &    0.33 &         0.10 \\
 $u - r$ &   -1.01 &    0.72 &         0.12 &   -0.49 &    0.45 &         0.13 &   -0.37 &    0.35 &         0.11 &   -0.41 &    0.31 &         0.10 &   -0.46 &    0.28 &         0.09 \\
 $u - i$ &   -1.17 &    0.67 &         0.10 &   -0.63 &    0.43 &         0.11 &   -0.48 &    0.33 &         0.10 &   -0.49 &    0.29 &         0.09 &   -0.53 &    0.26 &         0.09 \\
 $u - z$ &   -1.24 &    0.62 &         0.10 &   -0.71 &    0.41 &         0.11 &   -0.54 &    0.32 &         0.10 &   -0.53 &    0.27 &         0.09 &   -0.56 &    0.24 &         0.09 \\
 $g - r$ &   -1.12 &    2.35 &         0.11 &   -0.71 &    1.66 &         0.09 &   -0.54 &    1.29 &         0.08 &   -0.53 &    1.11 &         0.08 &   -0.55 &    0.97 &         0.08 \\
 $g - i$ &   -1.16 &    1.53 &         0.14 &   -0.76 &    1.10 &         0.09 &   -0.59 &    0.86 &         0.08 &   -0.56 &    0.73 &         0.08 &   -0.57 &    0.63 &         0.08 \\
 $g - z$ &   -1.15 &    1.17 &         0.16 &   -0.76 &    0.85 &         0.10 &   -0.58 &    0.67 &         0.09 &   -0.55 &    0.56 &         0.09 &   -0.56 &    0.48 &         0.09 \\
 $r - i$ &   -0.86 &    3.44 &         0.21 &   -0.59 &    2.60 &         0.14 &   -0.46 &    2.04 &         0.11 &   -0.44 &    1.70 &         0.11 &   -0.45 &    1.43 &         0.11 \\
 $r - z$ &   -0.84 &    1.88 &         0.22 &   -0.57 &    1.41 &         0.14 &   -0.44 &    1.11 &         0.12 &   -0.42 &    0.92 &         0.11 &   -0.43 &    0.76 &         0.11 \\
 $i - z$ &   -0.60 &    3.51 &         0.24 &   -0.40 &    2.67 &         0.16 &   -0.30 &    2.08 &         0.13 &   -0.30 &    1.70 &         0.12 &   -0.31 &    1.37 &         0.12 \\
\hline
\multicolumn{16}{c}{Sb galaxies}\\\hline
 $u - g$ &   -1.02 &    1.06 &         0.17 &   -0.56 &    0.69 &         0.17 &   -0.41 &    0.53 &         0.14 &   -0.43 &    0.45 &         0.13 &   -0.47 &    0.40 &         0.12 \\
 $u - r$ &   -1.19 &    0.80 &         0.13 &   -0.71 &    0.54 &         0.14 &   -0.53 &    0.42 &         0.12 &   -0.53 &    0.36 &         0.11 &   -0.56 &    0.31 &         0.10 \\
 $u - i$ &   -1.28 &    0.71 &         0.11 &   -0.80 &    0.49 &         0.12 &   -0.60 &    0.38 &         0.11 &   -0.59 &    0.32 &         0.10 &   -0.60 &    0.28 &         0.10 \\
 $u - z$ &   -1.33 &    0.64 &         0.10 &   -0.85 &    0.45 &         0.11 &   -0.64 &    0.35 &         0.10 &   -0.62 &    0.30 &         0.10 &   -0.63 &    0.26 &         0.10 \\
 $g - r$ &   -1.14 &    2.34 &         0.10 &   -0.75 &    1.71 &         0.09 &   -0.57 &    1.33 &         0.09 &   -0.56 &    1.13 &         0.08 &   -0.58 &    0.99 &         0.09 \\
 $g - i$ &   -1.21 &    1.54 &         0.11 &   -0.82 &    1.13 &         0.09 &   -0.63 &    0.88 &         0.09 &   -0.60 &    0.75 &         0.09 &   -0.62 &    0.65 &         0.09 \\
 $g - z$ &   -1.23 &    1.19 &         0.13 &   -0.84 &    0.89 &         0.10 &   -0.64 &    0.69 &         0.09 &   -0.62 &    0.59 &         0.09 &   -0.62 &    0.50 &         0.09 \\
 $r - i$ &   -1.11 &    3.83 &         0.18 &   -0.78 &    2.92 &         0.13 &   -0.60 &    2.29 &         0.11 &   -0.57 &    1.92 &         0.10 &   -0.58 &    1.63 &         0.10 \\
 $r - z$ &   -1.09 &    2.08 &         0.19 &   -0.76 &    1.59 &         0.13 &   -0.59 &    1.24 &         0.12 &   -0.56 &    1.04 &         0.11 &   -0.55 &    0.87 &         0.11 \\
 $i - z$ &   -0.91 &    4.07 &         0.22 &   -0.63 &    3.14 &         0.16 &   -0.48 &    2.43 &         0.13 &   -0.46 &    2.02 &         0.12 &   -0.47 &    1.66 &         0.12 \\
\hline
\multicolumn{16}{c}{Sbc galaxies}\\\hline
 $u - g$ &   -1.20 &    1.19 &         0.17 &   -0.75 &    0.83 &         0.17 &   -0.58 &    0.65 &         0.15 &   -0.58 &    0.57 &         0.13 &   -0.62 &    0.51 &         0.12 \\
 $u - r$ &   -1.31 &    0.85 &         0.12 &   -0.86 &    0.62 &         0.14 &   -0.67 &    0.49 &         0.12 &   -0.66 &    0.43 &         0.11 &   -0.69 &    0.38 &         0.10 \\
 $u - i$ &   -1.38 &    0.75 &         0.10 &   -0.94 &    0.55 &         0.12 &   -0.73 &    0.43 &         0.11 &   -0.71 &    0.38 &         0.10 &   -0.73 &    0.34 &         0.10 \\
 $u - z$ &   -1.43 &    0.68 &         0.10 &   -0.98 &    0.51 &         0.11 &   -0.77 &    0.40 &         0.10 &   -0.74 &    0.35 &         0.10 &   -0.76 &    0.31 &         0.10 \\
 $g - r$ &   -1.15 &    2.28 &         0.09 &   -0.81 &    1.78 &         0.09 &   -0.63 &    1.40 &         0.09 &   -0.62 &    1.22 &         0.08 &   -0.66 &    1.09 &         0.08 \\
 $g - i$ &   -1.23 &    1.50 &         0.11 &   -0.89 &    1.19 &         0.09 &   -0.70 &    0.94 &         0.09 &   -0.68 &    0.81 &         0.09 &   -0.70 &    0.72 &         0.09 \\
 $g - z$ &   -1.28 &    1.18 &         0.12 &   -0.93 &    0.94 &         0.10 &   -0.73 &    0.75 &         0.09 &   -0.71 &    0.64 &         0.09 &   -0.72 &    0.57 &         0.09 \\
 $r - i$ &   -1.23 &    3.90 &         0.17 &   -0.92 &    3.16 &         0.13 &   -0.72 &    2.51 &         0.11 &   -0.69 &    2.15 &         0.11 &   -0.70 &    1.88 &         0.11 \\
 $r - z$ &   -1.26 &    2.19 &         0.18 &   -0.94 &    1.77 &         0.14 &   -0.74 &    1.40 &         0.12 &   -0.70 &    1.20 &         0.11 &   -0.71 &    1.04 &         0.11 \\
 $i - z$ &   -1.17 &    4.53 &         0.21 &   -0.86 &    3.65 &         0.17 &   -0.67 &    2.87 &         0.14 &   -0.64 &    2.43 &         0.13 &   -0.64 &    2.08 &         0.13 \\
\hline
\multicolumn{16}{c}{Sc galaxies}\\\hline
 $u - g$ &   -1.15 &    1.06 &         0.20 &   -0.72 &    0.73 &         0.20 &   -0.55 &    0.56 &         0.17 &   -0.56 &    0.49 &         0.16 &   -0.61 &    0.44 &         0.15 \\
 $u - r$ &   -1.31 &    0.83 &         0.15 &   -0.89 &    0.61 &         0.16 &   -0.69 &    0.48 &         0.15 &   -0.68 &    0.41 &         0.14 &   -0.71 &    0.37 &         0.13 \\
 $u - i$ &   -1.41 &    0.74 &         0.13 &   -0.99 &    0.56 &         0.15 &   -0.77 &    0.44 &         0.14 &   -0.75 &    0.38 &         0.13 &   -0.77 &    0.34 &         0.12 \\
 $u - z$ &   -1.47 &    0.68 &         0.13 &   -1.04 &    0.52 &         0.14 &   -0.82 &    0.41 &         0.13 &   -0.79 &    0.35 &         0.12 &   -0.80 &    0.31 &         0.12 \\
 $g - r$ &   -1.18 &    2.24 &         0.13 &   -0.86 &    1.81 &         0.12 &   -0.68 &    1.44 &         0.11 &   -0.67 &    1.25 &         0.11 &   -0.70 &    1.12 &         0.11 \\
 $g - i$ &   -1.26 &    1.47 &         0.13 &   -0.95 &    1.21 &         0.12 &   -0.76 &    0.97 &         0.11 &   -0.73 &    0.83 &         0.11 &   -0.75 &    0.74 &         0.11 \\
 $g - z$ &   -1.31 &    1.15 &         0.15 &   -0.99 &    0.95 &         0.13 &   -0.79 &    0.76 &         0.12 &   -0.76 &    0.65 &         0.12 &   -0.77 &    0.58 &         0.12 \\
 $r - i$ &   -1.22 &    3.53 &         0.20 &   -0.93 &    2.97 &         0.16 &   -0.74 &    2.38 &         0.14 &   -0.71 &    2.03 &         0.13 &   -0.73 &    1.78 &         0.13 \\
 $r - z$ &   -1.25 &    1.98 &         0.21 &   -0.95 &    1.66 &         0.17 &   -0.76 &    1.32 &         0.15 &   -0.72 &    1.13 &         0.14 &   -0.73 &    0.98 &         0.14 \\
 $i - z$ &   -1.11 &    3.80 &         0.24 &   -0.83 &    3.15 &         0.20 &   -0.66 &    2.50 &         0.17 &   -0.63 &    2.10 &         0.16 &   -0.64 &    1.79 &         0.15 \\
\hline
\multicolumn{16}{c}{Sd galaxies}\\\hline
 $u - g$ &   -1.15 &    0.82 &         0.19 &   -0.73 &    0.51 &         0.18 &   -0.57 &    0.38 &         0.16 &   -0.57 &    0.32 &         0.14 &   -0.61 &    0.28 &         0.13 \\
 $u - r$ &   -1.28 &    0.71 &         0.15 &   -0.86 &    0.49 &         0.16 &   -0.68 &    0.37 &         0.14 &   -0.66 &    0.32 &         0.13 &   -0.69 &    0.28 &         0.12 \\
 $u - i$ &   -1.38 &    0.67 &         0.14 &   -0.94 &    0.47 &         0.15 &   -0.75 &    0.37 &         0.14 &   -0.72 &    0.31 &         0.12 &   -0.74 &    0.27 &         0.12 \\
 $u - z$ &   -1.45 &    0.63 &         0.13 &   -1.00 &    0.44 &         0.15 &   -0.79 &    0.35 &         0.13 &   -0.75 &    0.29 &         0.12 &   -0.77 &    0.25 &         0.12 \\
 $g - r$ &   -1.21 &    2.24 &         0.11 &   -0.89 &    1.78 &         0.12 &   -0.71 &    1.43 &         0.11 &   -0.69 &    1.22 &         0.10 &   -0.72 &    1.09 &         0.10 \\
 $g - i$ &   -1.33 &    1.52 &         0.12 &   -0.99 &    1.23 &         0.11 &   -0.80 &    0.99 &         0.11 &   -0.76 &    0.84 &         0.10 &   -0.78 &    0.74 &         0.10 \\
 $g - z$ &   -1.40 &    1.21 &         0.13 &   -1.05 &    0.98 &         0.12 &   -0.84 &    0.78 &         0.11 &   -0.79 &    0.66 &         0.11 &   -0.81 &    0.58 &         0.10 \\
 $r - i$ &   -1.32 &    3.58 &         0.17 &   -1.01 &    3.00 &         0.14 &   -0.81 &    2.41 &         0.12 &   -0.77 &    2.03 &         0.12 &   -0.78 &    1.76 &         0.11 \\
 $r - z$ &   -1.35 &    1.99 &         0.18 &   -1.02 &    1.62 &         0.15 &   -0.81 &    1.29 &         0.13 &   -0.77 &    1.08 &         0.12 &   -0.78 &    0.93 &         0.12 \\
 $i - z$ &   -1.21 &    3.50 &         0.21 &   -0.89 &    2.78 &         0.18 &   -0.70 &    2.18 &         0.15 &   -0.67 &    1.79 &         0.14 &   -0.68 &    1.48 &         0.13 \\
\hline
\footnote{Linear fitting at 0 $\leq$ R$_{HLR}$ $\leq$ 2 to optical log(M/L$_{\lambda}$)-color with base CBe and Chabrier IMF. The monochromatic mass-to-light ratios $(M/L_{\lambda})$ are in solar units. The SDSS $ugriz$ filters are in the AB magnitude system. $\sigma_{\lambda}$ is the scatter of the residuals of the relation log(M/L$_{\lambda}$)-color.}
\end{longtable}
}

\begin{table*}
\caption{Linear fitting at 0 $\leq$ R$_{HLR}$ $\leq$ 2 to $log(M/L_{\lambda}) = a_{\lambda} + (b_{\lambda} \times color)$ measured in restframe in the synthetic spectra (luminosities \textit{not} corrected for reddening).}
\label{tb:hubtyp_SynR_Mag_px1}
\centering
\begin{tabular}{lllllllllll}
\hline\hline
   Color & $a_{B}$ & $b_{B}$ & $\sigma_{B}$ & $a_{V}$ & $b_{V}$ & $\sigma_{V}$ & $a_{R}$ & $b_{R}$ & $\sigma_{R}$ \\
\hline

\multicolumn{10}{c}{E galaxies}\\\hline
 $B - V$ &   -1.81 &    2.69 &         0.09 &   -1.57 &    2.30 &         0.08 &   -1.44 &    2.03 &         0.08 \\
 $B - R$ &   -2.05 &    1.62 &         0.07 &   -1.78 &    1.40 &         0.07 &   -1.63 &    1.23 &         0.07 \\
 $V - R$ &   -2.12 &    3.69 &         0.08 &   -1.85 &    3.18 &         0.07 &   -1.68 &    2.80 &         0.07 \\
\hline
\multicolumn{10}{c}{S0 galaxies}\\\hline
 $B - V$ &   -1.25 &    2.07 &         0.10 &   -1.03 &    1.70 &         0.10 &   -0.94 &    1.47 &         0.09 \\
 $B - R$ &   -1.58 &    1.33 &         0.08 &   -1.32 &    1.11 &         0.08 &   -1.19 &    0.96 &         0.08 \\
 $V - R$ &   -1.81 &    3.25 &         0.09 &   -1.53 &    2.73 &         0.08 &   -1.38 &    2.38 &         0.08 \\
\hline
\multicolumn{10}{c}{Sa galaxies}\\\hline
 $B - V$ &   -0.92 &    1.73 &         0.09 &   -0.70 &    1.36 &         0.09 &   -0.63 &    1.15 &         0.08 \\
 $B - R$ &   -1.20 &    1.11 &         0.09 &   -0.93 &    0.88 &         0.08 &   -0.82 &    0.73 &         0.08 \\
 $V - R$ &   -1.28 &    2.47 &         0.12 &   -1.00 &    1.98 &         0.10 &   -0.86 &    1.63 &         0.10 \\
\hline
\multicolumn{10}{c}{Sb galaxies}\\\hline
 $B - V$ &   -1.02 &    1.84 &         0.10 &   -0.78 &    1.46 &         0.10 &   -0.69 &    1.21 &         0.09 \\
 $B - R$ &   -1.26 &    1.13 &         0.09 &   -0.98 &    0.90 &         0.09 &   -0.85 &    0.75 &         0.08 \\
 $V - R$ &   -1.47 &    2.68 &         0.11 &   -1.15 &    2.15 &         0.10 &   -0.99 &    1.78 &         0.09 \\
\hline
\multicolumn{10}{c}{Sbc galaxies}\\\hline
 $B - V$ &   -1.09 &    1.93 &         0.10 &   -0.86 &    1.56 &         0.09 &   -0.77 &    1.32 &         0.09 \\
 $B - R$ &   -1.34 &    1.18 &         0.09 &   -1.06 &    0.96 &         0.09 &   -0.94 &    0.81 &         0.08 \\
 $V - R$ &   -1.58 &    2.80 &         0.11 &   -1.27 &    2.29 &         0.10 &   -1.11 &    1.92 &         0.09 \\
\hline
\multicolumn{10}{c}{Sc galaxies}\\\hline
 $B - V$ &   -1.15 &    1.98 &         0.13 &   -0.92 &    1.61 &         0.13 &   -0.82 &    1.36 &         0.12 \\
 $B - R$ &   -1.42 &    1.22 &         0.12 &   -1.15 &    1.00 &         0.12 &   -1.02 &    0.84 &         0.11 \\
 $V - R$ &   -1.63 &    2.79 &         0.14 &   -1.33 &    2.31 &         0.12 &   -1.17 &    1.94 &         0.12 \\
\hline
\multicolumn{10}{c}{Sd galaxies}\\\hline
 $B - V$ &   -1.18 &    1.91 &         0.12 &   -0.96 &    1.58 &         0.12 &   -0.86 &    1.32 &         0.11 \\
 $B - R$ &   -1.47 &    1.23 &         0.11 &   -1.20 &    1.02 &         0.11 &   -1.06 &    0.86 &         0.10 \\
 $V - R$ &   -1.72 &    2.90 &         0.12 &   -1.42 &    2.42 &         0.11 &   -1.24 &    2.03 &         0.11 \\
\hline
\end{tabular}
\tablefoot{Linear fitting at 0 $\leq$ R$_{HLR}$ $\leq$ 2 to optical log(M/L$_{\lambda}$)-color with base CBe and Chabrier IMF. The monochromatic mass-to-light ratios $(M/L_{\lambda})$ are in solar units. The Johnson-Cousins $BVR$ filters are in the Vega magnitude system. $\sigma_{\lambda}$ is the scatter of the residuals of the relation log(M/L$_{\lambda}$)-color.}
\end{table*}

\begin{appendix}
\label{Apx}
\section{Intrinsic mass-to-light color relations}
In this appendix we include the intrinsic mass-to-light color relations, that is, the MLRC 
where the luminosities have been corrected for reddening. These relations can be useful 
for comparison with models or simulations.

\begin{table*}
\caption{Linear fitting at 0 $\leq$ R$_{HLR}$ $\leq$ 2 to $log(M/L_{\lambda}) = a_{\lambda} + (b_{\lambda} \times color)$ measured in restframe in the synthetic spectra and corrected for reddening (i.e. intrinsic luminosities).}
\label{tb:AVTABLE_SynRD_AB_px1}
\centering
\begin{tabular}{lllllllllllllllll}
\hline\hline
   Color & $a_{u}$ & $b_{u}$ & $\sigma_{u}$ & $a_{g}$ & $b_{g}$ & $\sigma_{g}$ & $a_{r}$ & $b_{r}$ & $\sigma_{r}$ & $a_{i}$ & $b_{i}$ & $\sigma_{i}$ & $a_{z}$ & $b_{z}$ & $\sigma_{z}$ \\
\hline

\multicolumn{16}{c}{All galaxies}\\\hline
 $u - g$ &   -1.26 &    1.15 &         0.18 &   -0.78 &    0.77 &         0.18 &   -0.59 &    0.60 &         0.16 &   -0.59 &    0.52 &         0.14 &   -0.61 &    0.47 &         0.13 \\
 $u - r$ &   -1.29 &    0.82 &         0.14 &   -0.82 &    0.56 &         0.15 &   -0.63 &    0.43 &         0.14 &   -0.62 &    0.38 &         0.12 &   -0.64 &    0.34 &         0.12 \\
 $u - i$ &   -1.36 &    0.73 &         0.12 &   -0.87 &    0.50 &         0.14 &   -0.67 &    0.39 &         0.13 &   -0.66 &    0.34 &         0.12 &   -0.68 &    0.31 &         0.11 \\
 $u - z$ &   -1.41 &    0.67 &         0.12 &   -0.91 &    0.46 &         0.13 &   -0.70 &    0.36 &         0.12 &   -0.68 &    0.32 &         0.11 &   -0.70 &    0.29 &         0.11 \\
 $g - r$ &   -1.25 &    2.58 &         0.10 &   -0.82 &    1.82 &         0.11 &   -0.63 &    1.43 &         0.10 &   -0.63 &    1.26 &         0.09 &   -0.65 &    1.14 &         0.09 \\
 $g - i$ &   -1.40 &    1.80 &         0.09 &   -0.93 &    1.28 &         0.09 &   -0.72 &    1.01 &         0.09 &   -0.70 &    0.89 &         0.09 &   -0.73 &    0.80 &         0.09 \\
 $g - z$ &   -1.48 &    1.47 &         0.10 &   -0.99 &    1.05 &         0.10 &   -0.77 &    0.82 &         0.09 &   -0.75 &    0.72 &         0.09 &   -0.76 &    0.66 &         0.09 \\
 $r - i$ &   -1.63 &    5.62 &         0.15 &   -1.11 &    4.02 &         0.11 &   -0.87 &    3.18 &         0.10 &   -0.83 &    2.80 &         0.09 &   -0.84 &    2.53 &         0.09 \\
 $r - z$ &   -1.69 &    3.24 &         0.16 &   -1.14 &    2.31 &         0.12 &   -0.89 &    1.82 &         0.11 &   -0.85 &    1.60 &         0.10 &   -0.86 &    1.44 &         0.10 \\
 $i - z$ &   -1.66 &    7.23 &         0.20 &   -1.11 &    5.11 &         0.16 &   -0.87 &    4.03 &         0.13 &   -0.83 &    3.53 &         0.12 &   -0.83 &    3.17 &         0.12 \\
\hline
\end{tabular}
\tablefoot{Linear fitting at 0 $\leq$ R$_{HLR}$ $\leq$ 2 to optical log(M/L$_{\lambda}$)-color with base CBe and Chabrier IMF. The monochromatic mass-to-light ratios $(M/L_{\lambda})$ are in solar units. The SDSS $ugriz$ filters are in the AB magnitude system. $\sigma_{\lambda}$ is the scatter of the residuals of the relation log(M/L$_{\lambda}$)-color.}
\end{table*}
\begin{table*}
\caption{Linear fitting at 0 $\leq$ R$_{HLR}$ $\leq$ 2 to $log(M/L_{\lambda}) = a_{\lambda} + (b_{\lambda} \times color)$ measured in restframe in the synthetic spectra and corrected for reddening (i.e. intrinsic luminosities).}
\label{tb:AVTABLE_SynRD_Mag_px1}
\centering
\begin{tabular}{lllllllllll}
\hline\hline
   Color & $a_{B}$ & $b_{B}$ & $\sigma_{B}$ & $a_{V}$ & $b_{V}$ & $\sigma_{V}$ & $a_{R}$ & $b_{R}$ & $\sigma_{R}$ \\
\hline

\multicolumn{10}{c}{All galaxies}\\\hline
 $B - V$ &   -1.08 &    1.89 &         0.12 &   -0.83 &    1.50 &         0.11 &   -0.75 &    1.28 &         0.11 \\
 $B - R$ &   -1.38 &    1.23 &         0.10 &   -1.07 &    0.98 &         0.10 &   -0.95 &    0.83 &         0.09 \\
 $V - R$ &   -1.85 &    3.34 &         0.10 &   -1.45 &    2.67 &         0.09 &   -1.28 &    2.28 &         0.09 \\
\hline
\end{tabular}
\tablefoot{Linear fitting at 0 $\leq$ R$_{HLR}$ $\leq$ 2 to optical log(M/L$_{\lambda}$)-color with base CBe and Chabrier IMF. The monochromatic mass-to-light ratios $(M/L_{\lambda})$ are in solar units. The Johnson-Cousins $BVR$ filters are in the Vega magnitude system. $\sigma_{\lambda}$ is the scatter of the residuals of the relation log(M/L$_{\lambda}$)-color.}
\end{table*}
\begin{table*}
\caption{Linear fitting at 0 $\leq$ R$_{HLR}$ $\leq$ 2 to $log(M/L_{\lambda}) = a_{\lambda} + (b_{\lambda} \times color)$ measured in restframe in the synthetic spectra and corrected for reddening (i.e. intrinsic luminosities).}
\label{tb:hub3_SynRD_AB_px1}
\centering
\begin{tabular}{lllllllllllllllll}
\hline\hline
   Color & $a_{u}$ & $b_{u}$ & $\sigma_{u}$ & $a_{g}$ & $b_{g}$ & $\sigma_{g}$ & $a_{r}$ & $b_{r}$ & $\sigma_{r}$ & $a_{i}$ & $b_{i}$ & $\sigma_{i}$ & $a_{z}$ & $b_{z}$ & $\sigma_{z}$ \\
\hline

\multicolumn{16}{c}{Early galaxies}\\\hline
 $u - g$ &   -0.94 &    0.95 &         0.14 &   -0.44 &    0.56 &         0.14 &   -0.30 &    0.42 &         0.12 &   -0.33 &    0.36 &         0.11 &   -0.38 &    0.32 &         0.11 \\
 $u - r$ &   -1.09 &    0.73 &         0.12 &   -0.56 &    0.44 &         0.12 &   -0.40 &    0.33 &         0.11 &   -0.42 &    0.29 &         0.11 &   -0.46 &    0.26 &         0.10 \\
 $u - i$ &   -1.22 &    0.67 &         0.11 &   -0.65 &    0.41 &         0.12 &   -0.48 &    0.31 &         0.11 &   -0.49 &    0.27 &         0.10 &   -0.52 &    0.24 &         0.10 \\
 $u - z$ &   -1.28 &    0.62 &         0.10 &   -0.70 &    0.39 &         0.11 &   -0.52 &    0.30 &         0.11 &   -0.52 &    0.26 &         0.10 &   -0.55 &    0.23 &         0.10 \\
 $g - r$ &   -1.32 &    2.68 &         0.09 &   -0.77 &    1.74 &         0.10 &   -0.59 &    1.35 &         0.09 &   -0.59 &    1.19 &         0.09 &   -0.62 &    1.07 &         0.08 \\
 $g - i$ &   -1.50 &    1.90 &         0.08 &   -0.92 &    1.25 &         0.08 &   -0.71 &    0.99 &         0.08 &   -0.69 &    0.87 &         0.08 &   -0.71 &    0.78 &         0.08 \\
 $g - z$ &   -1.54 &    1.52 &         0.08 &   -0.96 &    1.01 &         0.08 &   -0.75 &    0.80 &         0.08 &   -0.72 &    0.70 &         0.08 &   -0.73 &    0.63 &         0.08 \\
 $r - i$ &   -1.51 &    5.36 &         0.13 &   -1.00 &    3.73 &         0.09 &   -0.80 &    3.01 &         0.08 &   -0.77 &    2.64 &         0.08 &   -0.78 &    2.36 &         0.08 \\
 $r - z$ &   -1.50 &    3.00 &         0.14 &   -0.98 &    2.07 &         0.10 &   -0.78 &    1.66 &         0.09 &   -0.75 &    1.45 &         0.09 &   -0.75 &    1.29 &         0.09 \\
 $i - z$ &   -1.31 &    6.19 &         0.17 &   -0.82 &    4.17 &         0.13 &   -0.65 &    3.32 &         0.11 &   -0.63 &    2.89 &         0.10 &   -0.63 &    2.53 &         0.10 \\
\hline
\multicolumn{16}{c}{Intermediate galaxies}\\\hline
 $u - g$ &   -1.15 &    1.09 &         0.17 &   -0.67 &    0.71 &         0.16 &   -0.49 &    0.55 &         0.14 &   -0.50 &    0.48 &         0.13 &   -0.53 &    0.43 &         0.12 \\
 $u - r$ &   -1.23 &    0.80 &         0.13 &   -0.75 &    0.54 &         0.14 &   -0.56 &    0.42 &         0.12 &   -0.56 &    0.37 &         0.11 &   -0.59 &    0.33 &         0.11 \\
 $u - i$ &   -1.31 &    0.72 &         0.11 &   -0.82 &    0.49 &         0.13 &   -0.62 &    0.38 &         0.11 &   -0.61 &    0.34 &         0.11 &   -0.64 &    0.30 &         0.10 \\
 $u - z$ &   -1.37 &    0.66 &         0.11 &   -0.86 &    0.46 &         0.12 &   -0.66 &    0.36 &         0.11 &   -0.64 &    0.31 &         0.10 &   -0.67 &    0.28 &         0.10 \\
 $g - r$ &   -1.20 &    2.51 &         0.09 &   -0.79 &    1.80 &         0.10 &   -0.60 &    1.42 &         0.09 &   -0.60 &    1.25 &         0.08 &   -0.63 &    1.14 &         0.08 \\
 $g - i$ &   -1.36 &    1.75 &         0.09 &   -0.90 &    1.27 &         0.09 &   -0.70 &    1.00 &         0.08 &   -0.68 &    0.88 &         0.08 &   -0.70 &    0.80 &         0.08 \\
 $g - z$ &   -1.45 &    1.44 &         0.09 &   -0.97 &    1.04 &         0.09 &   -0.75 &    0.82 &         0.08 &   -0.73 &    0.73 &         0.08 &   -0.75 &    0.66 &         0.08 \\
 $r - i$ &   -1.54 &    5.28 &         0.13 &   -1.06 &    3.89 &         0.10 &   -0.83 &    3.09 &         0.09 &   -0.79 &    2.73 &         0.08 &   -0.81 &    2.47 &         0.08 \\
 $r - z$ &   -1.61 &    3.07 &         0.14 &   -1.10 &    2.25 &         0.11 &   -0.86 &    1.79 &         0.09 &   -0.82 &    1.58 &         0.09 &   -0.83 &    1.43 &         0.09 \\
 $i - z$ &   -1.52 &    6.58 &         0.18 &   -1.03 &    4.80 &         0.14 &   -0.81 &    3.81 &         0.12 &   -0.77 &    3.35 &         0.11 &   -0.78 &    3.01 &         0.11 \\
\hline
\multicolumn{16}{c}{Late galaxies}\\\hline
 $u - g$ &   -1.25 &    1.02 &         0.20 &   -0.81 &    0.69 &         0.20 &   -0.62 &    0.53 &         0.18 &   -0.62 &    0.47 &         0.16 &   -0.65 &    0.43 &         0.15 \\
 $u - r$ &   -1.34 &    0.82 &         0.16 &   -0.91 &    0.58 &         0.17 &   -0.71 &    0.46 &         0.16 &   -0.69 &    0.41 &         0.14 &   -0.72 &    0.38 &         0.14 \\
 $u - i$ &   -1.43 &    0.75 &         0.14 &   -0.99 &    0.55 &         0.16 &   -0.77 &    0.44 &         0.15 &   -0.75 &    0.39 &         0.13 &   -0.78 &    0.36 &         0.13 \\
 $u - z$ &   -1.50 &    0.71 &         0.14 &   -1.04 &    0.52 &         0.15 &   -0.82 &    0.41 &         0.14 &   -0.79 &    0.37 &         0.13 &   -0.81 &    0.34 &         0.13 \\
 $g - r$ &   -1.22 &    2.44 &         0.12 &   -0.87 &    1.91 &         0.13 &   -0.68 &    1.54 &         0.12 &   -0.67 &    1.37 &         0.11 &   -0.71 &    1.27 &         0.11 \\
 $g - i$ &   -1.36 &    1.73 &         0.12 &   -0.99 &    1.37 &         0.12 &   -0.79 &    1.11 &         0.11 &   -0.77 &    0.99 &         0.10 &   -0.79 &    0.91 &         0.10 \\
 $g - z$ &   -1.45 &    1.43 &         0.12 &   -1.07 &    1.13 &         0.12 &   -0.85 &    0.92 &         0.11 &   -0.82 &    0.81 &         0.11 &   -0.83 &    0.75 &         0.10 \\
 $r - i$ &   -1.46 &    4.70 &         0.18 &   -1.10 &    3.85 &         0.15 &   -0.88 &    3.14 &         0.13 &   -0.84 &    2.77 &         0.12 &   -0.86 &    2.55 &         0.12 \\
 $r - z$ &   -1.53 &    2.78 &         0.19 &   -1.15 &    2.24 &         0.15 &   -0.92 &    1.82 &         0.13 &   -0.87 &    1.60 &         0.13 &   -0.88 &    1.46 &         0.12 \\
 $i - z$ &   -1.39 &    5.33 &         0.23 &   -1.01 &    4.17 &         0.19 &   -0.80 &    3.36 &         0.16 &   -0.76 &    2.93 &         0.15 &   -0.78 &    2.65 &         0.15 \\
\hline
\end{tabular}
\tablefoot{Linear fitting at 0 $\leq$ R$_{HLR}$ $\leq$ 2 to optical log(M/L$_{\lambda}$)-color with base CBe and Chabrier IMF. The monochromatic mass-to-light ratios $(M/L_{\lambda})$ are in solar units. The SDSS $ugriz$ filters are in the AB magnitude system. $\sigma_{\lambda}$ is the scatter of the residuals of the relation log(M/L$_{\lambda}$)-color.}
\end{table*}
\begin{table*}
\caption{Linear fitting at 0 $\leq$ R$_{HLR}$ $\leq$ 2 to $log(M/L_{\lambda}) = a_{\lambda} + (b_{\lambda} \times color)$ measured in restframe in the synthetic spectra and corrected for reddening (i.e. intrinsic luminosities).}
\label{tb:hub3_SynRD_Mag_px1}
\centering
\begin{tabular}{lllllllllll}
\hline\hline
   Color & $a_{B}$ & $b_{B}$ & $\sigma_{B}$ & $a_{V}$ & $b_{V}$ & $\sigma_{V}$ & $a_{R}$ & $b_{R}$ & $\sigma_{R}$ \\
\hline

\multicolumn{10}{c}{Early galaxies}\\\hline
 $B - V$ &   -0.95 &    1.72 &         0.10 &   -0.70 &    1.33 &         0.10 &   -0.64 &    1.13 &         0.10 \\
 $B - R$ &   -1.31 &    1.17 &         0.09 &   -1.00 &    0.92 &         0.09 &   -0.89 &    0.78 &         0.09 \\
 $V - R$ &   -1.84 &    3.33 &         0.08 &   -1.45 &    2.66 &         0.08 &   -1.28 &    2.28 &         0.08 \\
\hline
\multicolumn{10}{c}{Intermediate galaxies}\\\hline
 $B - V$ &   -1.03 &    1.87 &         0.11 &   -0.79 &    1.48 &         0.10 &   -0.71 &    1.26 &         0.09 \\
 $B - R$ &   -1.34 &    1.22 &         0.09 &   -1.04 &    0.97 &         0.09 &   -0.92 &    0.83 &         0.08 \\
 $V - R$ &   -1.81 &    3.29 &         0.09 &   -1.42 &    2.64 &         0.09 &   -1.25 &    2.26 &         0.08 \\
\hline
\multicolumn{10}{c}{Late galaxies}\\\hline
 $B - V$ &   -1.18 &    2.05 &         0.14 &   -0.94 &    1.69 &         0.13 &   -0.85 &    1.45 &         0.12 \\
 $B - R$ &   -1.50 &    1.34 &         0.12 &   -1.21 &    1.11 &         0.12 &   -1.08 &    0.96 &         0.11 \\
 $V - R$ &   -1.85 &    3.31 &         0.13 &   -1.51 &    2.76 &         0.12 &   -1.34 &    2.38 &         0.11 \\
\hline
\end{tabular}
\tablefoot{Linear fitting at 0 $\leq$ R$_{HLR}$ $\leq$ 2 to optical log(M/L$_{\lambda}$)-color with base CBe and Chabrier IMF. The monochromatic mass-to-light ratios $(M/L_{\lambda})$ are in solar units. The Johnson-Cousins $BVR$ filters are in the Vega magnitude system. $\sigma_{\lambda}$ is the scatter of the residuals of the relation log(M/L$_{\lambda}$)-color.}
\end{table*}
\longtab[5]{
\begin{longtable}{llllllllllllllll}
\caption{\label{tb:hubtyp_SynRD_AB_px1} Linear fitting at 0 $\leq$ R$_{HLR}$ $\leq$ 2 to $log(M/L_{\lambda}) = a_{\lambda} + (b_{\lambda} \times color)$ measured in restframe in the synthetic spectra and corrected for reddening (i.e. intrinsic luminosities).}\\
\hline\hline
   Color & $a_{u}$ & $b_{u}$ & $\sigma_{u}$ & $a_{g}$ & $b_{g}$ & $\sigma_{g}$ & $a_{r}$ & $b_{r}$ & $\sigma_{r}$ & $a_{i}$ & $b_{i}$ & $\sigma_{i}$ & $a_{z}$ & $b_{z}$ & $\sigma_{z}$ \\
\hline
\endfirsthead
\caption{continued.}\\
\hline\hline
   Color & $a_{u}$ & $b_{u}$ & $\sigma_{u}$ & $a_{g}$ & $b_{g}$ & $\sigma_{g}$ & $a_{r}$ & $b_{r}$ & $\sigma_{r}$ & $a_{i}$ & $b_{i}$ & $\sigma_{i}$ & $a_{z}$ & $b_{z}$ & $\sigma_{z}$ \\
\endhead
\hline
\endfoot
\multicolumn{16}{c}{E galaxies}\\\hline
 $u - g$ &   -1.54 &    1.30 &         0.13 &   -1.01 &    0.89 &         0.13 &   -0.85 &    0.73 &         0.12 &   -0.81 &    0.64 &         0.11 &   -0.80 &    0.56 &         0.11 \\
 $u - r$ &   -1.78 &    1.00 &         0.11 &   -1.24 &    0.71 &         0.11 &   -1.06 &    0.60 &         0.11 &   -1.00 &    0.53 &         0.10 &   -0.98 &    0.47 &         0.10 \\
 $u - i$ &   -1.84 &    0.88 &         0.10 &   -1.31 &    0.64 &         0.10 &   -1.13 &    0.54 &         0.10 &   -1.07 &    0.48 &         0.09 &   -1.05 &    0.43 &         0.09 \\
 $u - z$ &   -1.82 &    0.79 &         0.09 &   -1.31 &    0.58 &         0.10 &   -1.13 &    0.49 &         0.10 &   -1.07 &    0.43 &         0.09 &   -1.05 &    0.39 &         0.09 \\
 $g - r$ &   -1.85 &    3.41 &         0.08 &   -1.44 &    2.63 &         0.08 &   -1.26 &    2.24 &         0.08 &   -1.20 &    2.01 &         0.08 &   -1.19 &    1.83 &         0.08 \\
 $g - i$ &   -1.77 &    2.14 &         0.06 &   -1.39 &    1.66 &         0.07 &   -1.23 &    1.43 &         0.07 &   -1.17 &    1.27 &         0.07 &   -1.16 &    1.16 &         0.07 \\
 $g - z$ &   -1.73 &    1.65 &         0.07 &   -1.35 &    1.27 &         0.07 &   -1.18 &    1.09 &         0.07 &   -1.13 &    0.97 &         0.07 &   -1.12 &    0.88 &         0.07 \\
 $r - i$ &   -1.44 &    5.24 &         0.08 &   -1.13 &    4.05 &         0.07 &   -1.00 &    3.48 &         0.07 &   -0.97 &    3.11 &         0.07 &   -0.97 &    2.82 &         0.07 \\
 $r - z$ &   -1.43 &    2.92 &         0.09 &   -1.11 &    2.24 &         0.08 &   -0.98 &    1.93 &         0.08 &   -0.95 &    1.71 &         0.08 &   -0.95 &    1.55 &         0.08 \\
 $i - z$ &   -1.23 &    5.99 &         0.12 &   -0.92 &    4.50 &         0.11 &   -0.81 &    3.83 &         0.10 &   -0.79 &    3.38 &         0.09 &   -0.79 &    3.02 &         0.09 \\
\hline
\multicolumn{16}{c}{S0 galaxies}\\\hline
 $u - g$ &   -1.04 &    1.01 &         0.14 &   -0.55 &    0.62 &         0.13 &   -0.41 &    0.47 &         0.12 &   -0.42 &    0.41 &         0.11 &   -0.45 &    0.36 &         0.11 \\
 $u - r$ &   -1.22 &    0.78 &         0.12 &   -0.70 &    0.50 &         0.12 &   -0.54 &    0.39 &         0.11 &   -0.54 &    0.34 &         0.10 &   -0.56 &    0.30 &         0.10 \\
 $u - i$ &   -1.33 &    0.71 &         0.10 &   -0.80 &    0.46 &         0.11 &   -0.63 &    0.36 &         0.11 &   -0.61 &    0.32 &         0.10 &   -0.63 &    0.28 &         0.10 \\
 $u - z$ &   -1.38 &    0.65 &         0.10 &   -0.85 &    0.43 &         0.11 &   -0.67 &    0.34 &         0.10 &   -0.65 &    0.30 &         0.10 &   -0.66 &    0.26 &         0.09 \\
 $g - r$ &   -1.36 &    2.74 &         0.09 &   -0.91 &    1.91 &         0.09 &   -0.73 &    1.54 &         0.09 &   -0.71 &    1.35 &         0.08 &   -0.72 &    1.21 &         0.08 \\
 $g - i$ &   -1.52 &    1.91 &         0.07 &   -1.01 &    1.33 &         0.08 &   -0.83 &    1.08 &         0.08 &   -0.80 &    0.96 &         0.07 &   -0.81 &    0.86 &         0.07 \\
 $g - z$ &   -1.58 &    1.54 &         0.07 &   -1.04 &    1.07 &         0.08 &   -0.85 &    0.86 &         0.08 &   -0.82 &    0.76 &         0.07 &   -0.82 &    0.68 &         0.07 \\
 $r - i$ &   -1.49 &    5.33 &         0.10 &   -1.06 &    3.88 &         0.08 &   -0.88 &    3.18 &         0.07 &   -0.84 &    2.81 &         0.07 &   -0.85 &    2.53 &         0.07 \\
 $r - z$ &   -1.50 &    3.01 &         0.11 &   -1.06 &    2.18 &         0.09 &   -0.87 &    1.78 &         0.08 &   -0.84 &    1.57 &         0.08 &   -0.83 &    1.40 &         0.08 \\
 $i - z$ &   -1.33 &    6.30 &         0.14 &   -0.91 &    4.49 &         0.11 &   -0.75 &    3.65 &         0.10 &   -0.72 &    3.19 &         0.09 &   -0.72 &    2.83 &         0.09 \\
\hline
\multicolumn{16}{c}{Sa galaxies}\\\hline
 $u - g$ &   -0.90 &    0.94 &         0.14 &   -0.40 &    0.55 &         0.14 &   -0.27 &    0.41 &         0.12 &   -0.30 &    0.36 &         0.11 &   -0.35 &    0.32 &         0.10 \\
 $u - r$ &   -1.04 &    0.72 &         0.12 &   -0.51 &    0.43 &         0.12 &   -0.35 &    0.32 &         0.11 &   -0.38 &    0.29 &         0.10 &   -0.43 &    0.26 &         0.09 \\
 $u - i$ &   -1.16 &    0.66 &         0.11 &   -0.59 &    0.40 &         0.11 &   -0.42 &    0.30 &         0.10 &   -0.44 &    0.27 &         0.09 &   -0.48 &    0.24 &         0.09 \\
 $u - z$ &   -1.23 &    0.61 &         0.10 &   -0.64 &    0.38 &         0.11 &   -0.46 &    0.29 &         0.10 &   -0.47 &    0.25 &         0.09 &   -0.51 &    0.22 &         0.09 \\
 $g - r$ &   -1.21 &    2.52 &         0.10 &   -0.68 &    1.63 &         0.09 &   -0.49 &    1.24 &         0.09 &   -0.51 &    1.11 &         0.08 &   -0.55 &    1.01 &         0.08 \\
 $g - i$ &   -1.40 &    1.81 &         0.09 &   -0.83 &    1.19 &         0.08 &   -0.62 &    0.92 &         0.08 &   -0.62 &    0.81 &         0.07 &   -0.64 &    0.73 &         0.07 \\
 $g - z$ &   -1.47 &    1.46 &         0.10 &   -0.89 &    0.97 &         0.08 &   -0.67 &    0.76 &         0.08 &   -0.65 &    0.66 &         0.08 &   -0.67 &    0.59 &         0.08 \\
 $r - i$ &   -1.46 &    5.14 &         0.17 &   -0.91 &    3.50 &         0.11 &   -0.70 &    2.77 &         0.09 &   -0.68 &    2.42 &         0.09 &   -0.69 &    2.16 &         0.09 \\
 $r - z$ &   -1.45 &    2.88 &         0.18 &   -0.89 &    1.95 &         0.12 &   -0.69 &    1.53 &         0.10 &   -0.67 &    1.34 &         0.10 &   -0.67 &    1.18 &         0.10 \\
 $i - z$ &   -1.23 &    5.77 &         0.21 &   -0.72 &    3.80 &         0.15 &   -0.54 &    2.96 &         0.12 &   -0.53 &    2.57 &         0.11 &   -0.54 &    2.24 &         0.11 \\
\hline
\multicolumn{16}{c}{Sb galaxies}\\\hline
 $u - g$ &   -1.04 &    1.02 &         0.16 &   -0.57 &    0.65 &         0.16 &   -0.41 &    0.49 &         0.14 &   -0.42 &    0.42 &         0.12 &   -0.46 &    0.38 &         0.12 \\
 $u - r$ &   -1.15 &    0.76 &         0.13 &   -0.66 &    0.49 &         0.13 &   -0.48 &    0.38 &         0.12 &   -0.48 &    0.33 &         0.11 &   -0.52 &    0.30 &         0.11 \\
 $u - i$ &   -1.24 &    0.68 &         0.11 &   -0.73 &    0.45 &         0.12 &   -0.54 &    0.34 &         0.11 &   -0.53 &    0.30 &         0.10 &   -0.57 &    0.27 &         0.10 \\
 $u - z$ &   -1.30 &    0.64 &         0.11 &   -0.78 &    0.42 &         0.12 &   -0.57 &    0.32 &         0.11 &   -0.57 &    0.28 &         0.10 &   -0.59 &    0.25 &         0.10 \\
 $g - r$ &   -1.19 &    2.49 &         0.10 &   -0.74 &    1.73 &         0.10 &   -0.56 &    1.34 &         0.09 &   -0.55 &    1.17 &         0.09 &   -0.59 &    1.06 &         0.08 \\
 $g - i$ &   -1.34 &    1.74 &         0.09 &   -0.86 &    1.22 &         0.09 &   -0.65 &    0.95 &         0.09 &   -0.64 &    0.83 &         0.08 &   -0.66 &    0.75 &         0.08 \\
 $g - z$ &   -1.44 &    1.44 &         0.09 &   -0.93 &    1.01 &         0.09 &   -0.71 &    0.78 &         0.08 &   -0.69 &    0.69 &         0.08 &   -0.70 &    0.62 &         0.08 \\
 $r - i$ &   -1.52 &    5.26 &         0.12 &   -1.01 &    3.76 &         0.10 &   -0.78 &    2.95 &         0.09 &   -0.75 &    2.59 &         0.08 &   -0.76 &    2.33 &         0.08 \\
 $r - z$ &   -1.60 &    3.08 &         0.12 &   -1.06 &    2.18 &         0.10 &   -0.82 &    1.71 &         0.09 &   -0.78 &    1.50 &         0.08 &   -0.79 &    1.35 &         0.08 \\
 $i - z$ &   -1.51 &    6.60 &         0.17 &   -0.99 &    4.65 &         0.13 &   -0.76 &    3.65 &         0.11 &   -0.73 &    3.19 &         0.10 &   -0.73 &    2.85 &         0.10 \\
\hline
\multicolumn{16}{c}{Sbc galaxies}\\\hline
 $u - g$ &   -1.19 &    1.11 &         0.17 &   -0.72 &    0.74 &         0.17 &   -0.54 &    0.58 &         0.15 &   -0.54 &    0.51 &         0.13 &   -0.58 &    0.47 &         0.13 \\
 $u - r$ &   -1.28 &    0.83 &         0.13 &   -0.82 &    0.58 &         0.14 &   -0.62 &    0.46 &         0.12 &   -0.62 &    0.40 &         0.11 &   -0.65 &    0.37 &         0.11 \\
 $u - i$ &   -1.37 &    0.74 &         0.11 &   -0.89 &    0.53 &         0.13 &   -0.69 &    0.42 &         0.11 &   -0.68 &    0.37 &         0.11 &   -0.70 &    0.34 &         0.10 \\
 $u - z$ &   -1.43 &    0.69 &         0.10 &   -0.94 &    0.50 &         0.12 &   -0.73 &    0.39 &         0.11 &   -0.71 &    0.35 &         0.10 &   -0.74 &    0.32 &         0.10 \\
 $g - r$ &   -1.20 &    2.48 &         0.08 &   -0.82 &    1.88 &         0.09 &   -0.64 &    1.50 &         0.08 &   -0.64 &    1.33 &         0.08 &   -0.67 &    1.23 &         0.08 \\
 $g - i$ &   -1.35 &    1.74 &         0.08 &   -0.94 &    1.32 &         0.08 &   -0.74 &    1.06 &         0.08 &   -0.72 &    0.94 &         0.07 &   -0.75 &    0.87 &         0.07 \\
 $g - z$ &   -1.44 &    1.43 &         0.09 &   -1.02 &    1.09 &         0.08 &   -0.80 &    0.87 &         0.08 &   -0.77 &    0.78 &         0.07 &   -0.79 &    0.71 &         0.07 \\
 $r - i$ &   -1.51 &    5.10 &         0.13 &   -1.09 &    3.99 &         0.11 &   -0.86 &    3.22 &         0.09 &   -0.83 &    2.86 &         0.09 &   -0.85 &    2.62 &         0.08 \\
 $r - z$ &   -1.57 &    2.97 &         0.15 &   -1.14 &    2.32 &         0.12 &   -0.90 &    1.87 &         0.10 &   -0.87 &    1.66 &         0.09 &   -0.88 &    1.52 &         0.09 \\
 $i - z$ &   -1.45 &    6.14 &         0.19 &   -1.04 &    4.77 &         0.15 &   -0.82 &    3.87 &         0.13 &   -0.79 &    3.42 &         0.12 &   -0.80 &    3.10 &         0.11 \\
\hline
\multicolumn{16}{c}{Sc galaxies}\\\hline
 $u - g$ &   -1.13 &    0.93 &         0.19 &   -0.68 &    0.59 &         0.19 &   -0.51 &    0.45 &         0.17 &   -0.52 &    0.40 &         0.15 &   -0.57 &    0.37 &         0.15 \\
 $u - r$ &   -1.26 &    0.77 &         0.15 &   -0.81 &    0.53 &         0.16 &   -0.62 &    0.41 &         0.15 &   -0.62 &    0.37 &         0.14 &   -0.66 &    0.35 &         0.13 \\
 $u - i$ &   -1.37 &    0.72 &         0.13 &   -0.91 &    0.51 &         0.15 &   -0.70 &    0.40 &         0.14 &   -0.70 &    0.36 &         0.13 &   -0.73 &    0.34 &         0.13 \\
 $u - z$ &   -1.44 &    0.68 &         0.13 &   -0.97 &    0.49 &         0.15 &   -0.75 &    0.39 &         0.14 &   -0.74 &    0.35 &         0.13 &   -0.76 &    0.32 &         0.12 \\
 $g - r$ &   -1.19 &    2.39 &         0.12 &   -0.83 &    1.83 &         0.12 &   -0.65 &    1.47 &         0.12 &   -0.65 &    1.32 &         0.11 &   -0.68 &    1.23 &         0.11 \\
 $g - i$ &   -1.35 &    1.72 &         0.11 &   -0.97 &    1.35 &         0.11 &   -0.77 &    1.09 &         0.11 &   -0.75 &    0.98 &         0.10 &   -0.78 &    0.91 &         0.10 \\
 $g - z$ &   -1.44 &    1.42 &         0.12 &   -1.05 &    1.12 &         0.11 &   -0.83 &    0.91 &         0.11 &   -0.81 &    0.81 &         0.10 &   -0.83 &    0.75 &         0.10 \\
 $r - i$ &   -1.40 &    4.53 &         0.18 &   -1.06 &    3.73 &         0.14 &   -0.85 &    3.06 &         0.13 &   -0.82 &    2.72 &         0.12 &   -0.84 &    2.50 &         0.12 \\
 $r - z$ &   -1.46 &    2.66 &         0.18 &   -1.09 &    2.15 &         0.15 &   -0.88 &    1.76 &         0.13 &   -0.84 &    1.56 &         0.12 &   -0.85 &    1.43 &         0.12 \\
 $i - z$ &   -1.25 &    4.76 &         0.22 &   -0.88 &    3.70 &         0.18 &   -0.70 &    3.00 &         0.15 &   -0.68 &    2.63 &         0.14 &   -0.70 &    2.39 &         0.14 \\
\hline
\multicolumn{16}{c}{Sd galaxies}\\\hline
 $u - g$ &   -1.23 &    0.83 &         0.19 &   -0.79 &    0.49 &         0.19 &   -0.61 &    0.36 &         0.16 &   -0.60 &    0.31 &         0.15 &   -0.63 &    0.28 &         0.14 \\
 $u - r$ &   -1.31 &    0.70 &         0.16 &   -0.87 &    0.45 &         0.16 &   -0.67 &    0.34 &         0.15 &   -0.66 &    0.30 &         0.13 &   -0.69 &    0.27 &         0.13 \\
 $u - i$ &   -1.39 &    0.66 &         0.14 &   -0.94 &    0.44 &         0.15 &   -0.72 &    0.34 &         0.14 &   -0.70 &    0.29 &         0.13 &   -0.73 &    0.26 &         0.12 \\
 $u - z$ &   -1.45 &    0.63 &         0.14 &   -0.98 &    0.42 &         0.15 &   -0.76 &    0.32 &         0.14 &   -0.73 &    0.28 &         0.13 &   -0.75 &    0.25 &         0.12 \\
 $g - r$ &   -1.21 &    2.31 &         0.13 &   -0.86 &    1.74 &         0.12 &   -0.67 &    1.36 &         0.12 &   -0.66 &    1.20 &         0.11 &   -0.69 &    1.10 &         0.10 \\
 $g - i$ &   -1.33 &    1.62 &         0.13 &   -0.97 &    1.25 &         0.12 &   -0.76 &    0.98 &         0.11 &   -0.73 &    0.85 &         0.10 &   -0.75 &    0.78 &         0.10 \\
 $g - z$ &   -1.41 &    1.33 &         0.13 &   -1.02 &    1.01 &         0.12 &   -0.80 &    0.79 &         0.11 &   -0.77 &    0.68 &         0.11 &   -0.78 &    0.62 &         0.10 \\
 $r - i$ &   -1.40 &    4.11 &         0.18 &   -1.03 &    3.22 &         0.14 &   -0.81 &    2.55 &         0.13 &   -0.77 &    2.19 &         0.12 &   -0.79 &    1.98 &         0.11 \\
 $r - z$ &   -1.46 &    2.38 &         0.19 &   -1.05 &    1.82 &         0.15 &   -0.83 &    1.42 &         0.13 &   -0.78 &    1.22 &         0.12 &   -0.80 &    1.09 &         0.12 \\
 $i - z$ &   -1.32 &    4.19 &         0.22 &   -0.92 &    3.05 &         0.18 &   -0.72 &    2.36 &         0.15 &   -0.69 &    2.00 &         0.14 &   -0.71 &    1.75 &         0.13 \\
\hline
\footnote{Linear fitting at 0 $\leq$ R$_{HLR}$ $\leq$ 2 to optical log(M/L$_{\lambda}$)-color with base CBe and Chabrier IMF. The monochromatic mass-to-light ratios $(M/L_{\lambda})$ are in solar units. The SDSS $ugriz$ filters are in the AB magnitude system. $\sigma_{\lambda}$ is the scatter of the residuals of the relation log(M/L$_{\lambda}$)-color.}
\end{longtable}
}

\begin{table*}
\caption{Linear fitting at 0 $\leq$ R$_{HLR}$ $\leq$ 2 to $log(M/L_{\lambda}) = a_{\lambda} + (b_{\lambda} \times color)$ measured in restframe in the synthetic spectra and corrected for reddening (i.e. intrinsic luminosities).}
\label{tb:hubtyp_SynRD_Mag_px1}
\centering
\begin{tabular}{lllllllllll}
\hline\hline
   Color & $a_{B}$ & $b_{B}$ & $\sigma_{B}$ & $a_{V}$ & $b_{V}$ & $\sigma_{V}$ & $a_{R}$ & $b_{R}$ & $\sigma_{R}$ \\
\hline

\multicolumn{10}{c}{E galaxies}\\\hline
 $B - V$ &   -1.73 &    2.60 &         0.09 &   -1.50 &    2.23 &         0.09 &   -1.38 &    1.97 &         0.09 \\
 $B - R$ &   -2.03 &    1.62 &         0.08 &   -1.77 &    1.40 &         0.08 &   -1.63 &    1.24 &         0.08 \\
 $V - R$ &   -2.29 &    3.95 &         0.07 &   -2.01 &    3.42 &         0.07 &   -1.84 &    3.04 &         0.07 \\
\hline
\multicolumn{10}{c}{S0 galaxies}\\\hline
 $B - V$ &   -1.11 &    1.90 &         0.10 &   -0.87 &    1.51 &         0.10 &   -0.79 &    1.30 &         0.09 \\
 $B - R$ &   -1.45 &    1.26 &         0.09 &   -1.17 &    1.02 &         0.09 &   -1.05 &    0.88 &         0.08 \\
 $V - R$ &   -1.93 &    3.45 &         0.07 &   -1.56 &    2.81 &         0.07 &   -1.40 &    2.44 &         0.07 \\
\hline
\multicolumn{10}{c}{Sa galaxies}\\\hline
 $B - V$ &   -0.86 &    1.64 &         0.10 &   -0.61 &    1.25 &         0.09 &   -0.56 &    1.07 &         0.09 \\
 $B - R$ &   -1.21 &    1.12 &         0.09 &   -0.90 &    0.87 &         0.08 &   -0.80 &    0.74 &         0.08 \\
 $V - R$ &   -1.71 &    3.15 &         0.09 &   -1.30 &    2.47 &         0.08 &   -1.15 &    2.11 &         0.08 \\
\hline
\multicolumn{10}{c}{Sb galaxies}\\\hline
 $B - V$ &   -0.96 &    1.77 &         0.11 &   -0.71 &    1.37 &         0.10 &   -0.64 &    1.16 &         0.10 \\
 $B - R$ &   -1.27 &    1.16 &         0.10 &   -0.96 &    0.91 &         0.09 &   -0.85 &    0.77 &         0.09 \\
 $V - R$ &   -1.73 &    3.17 &         0.10 &   -1.33 &    2.51 &         0.09 &   -1.17 &    2.13 &         0.08 \\
\hline
\multicolumn{10}{c}{Sbc galaxies}\\\hline
 $B - V$ &   -1.08 &    1.95 &         0.10 &   -0.84 &    1.57 &         0.10 &   -0.76 &    1.36 &         0.09 \\
 $B - R$ &   -1.41 &    1.27 &         0.09 &   -1.11 &    1.03 &         0.08 &   -0.99 &    0.89 &         0.08 \\
 $V - R$ &   -1.86 &    3.40 &         0.09 &   -1.49 &    2.77 &         0.08 &   -1.32 &    2.39 &         0.08 \\
\hline
\multicolumn{10}{c}{Sc galaxies}\\\hline
 $B - V$ &   -1.10 &    1.91 &         0.13 &   -0.86 &    1.56 &         0.13 &   -0.78 &    1.35 &         0.12 \\
 $B - R$ &   -1.44 &    1.29 &         0.12 &   -1.15 &    1.06 &         0.11 &   -1.04 &    0.92 &         0.11 \\
 $V - R$ &   -1.81 &    3.26 &         0.12 &   -1.47 &    2.71 &         0.12 &   -1.31 &    2.35 &         0.11 \\
\hline
\multicolumn{10}{c}{Sd galaxies}\\\hline
 $B - V$ &   -1.17 &    1.91 &         0.13 &   -0.93 &    1.55 &         0.13 &   -0.83 &    1.31 &         0.12 \\
 $B - R$ &   -1.45 &    1.24 &         0.12 &   -1.16 &    1.01 &         0.12 &   -1.03 &    0.86 &         0.11 \\
 $V - R$ &   -1.72 &    2.95 &         0.13 &   -1.39 &    2.41 &         0.12 &   -1.21 &    2.03 &         0.11 \\
\hline
\end{tabular}
\tablefoot{Linear fitting at 0 $\leq$ R$_{HLR}$ $\leq$ 2 to optical log(M/L$_{\lambda}$)-color with base CBe and Chabrier IMF. The monochromatic mass-to-light ratios $(M/L_{\lambda})$ are in solar units. The Johnson-Cousins $BVR$ filters are in the Vega magnitude system. $\sigma_{\lambda}$ is the scatter of the residuals of the relation log(M/L$_{\lambda}$)-color.}
\end{table*}
\end{appendix}

\end{document}